# Alloying driven transition between ferro- and antiferromagnetism in UTGe compounds: the $UCo_{1-x}Ir_xGe$ case

Dávid Hovančík[1,2,], Akinari Koriki[1,3], Anežka Bendová[1], Petr Doležal[1], Petr Proschek[1], Martin Míšek[4], Marian Reiffers[2], Jan Prokleška[1], Jiří Pospíšil[1] and Vladimír Sechovský[1]

*[1]Charles University, Faculty of Mathematics and Physics, Department of Condensed Matter Physics, Ke Karlovu 5, 121 16 Prague 2, Czech Republic*
*[2]University of Presov, Faculty of Humanities and Natural Sciences, 17 Novembra 1, 081 16 Presov, Slovakia*
*[3]Hokkaido University, Graduate School of Science, Department of Condensed Matter Physics, Kita10, Nishi 8, Kita-ku, Sapporo, 060-0810, Japan*
*[4]Institute of Physics, Academy of Sciences of Czech Republic, v.v.i, Na Slovance 2, 182 21 Prague 8, Czech Republic*

## Abstract

The evolution of magnetic properties of isostructural and isoelectronic solid solutions of the superconducting itinerant 5*f*-electron ferromagnet UCoGe with antiferromagnet UIrGe was studied by magnetization, AC susceptibility, specific heat, and electrical resistivity measurements of a series of $UCo_{1-x}Ir_xGe$ compounds in polycrystalline and single crystalline form at various temperatures and magnetic fields. Both the weak ferromagnetism and superconductivity in UCoGe were found to have vanished already for very low Ir substitution for Co ($x$ = 0.02). The antiferromagnetism of UIrGe is gradually suppressed. This is documented by a rapid decrease in both Néel temperature and the critical field of the metamagnetic transition with decreasing Ir concentration, which both tend to vanish just above $x = 0.8$. The section of the *T*-*x* phase diagram in the range $0.02 \leq x \leq 0.8$ is dominated by a correlated paramagnetic phase exhibiting very broad bumps in temperature dependencies of *b*-axis magnetization and specific heat developing with increasing $x$. For $x \geq 0.24$, wide peaks appear in the *c*-axis thermomagnetic curves due to antiferromagnetic correlations which may eventually lead to frozen incoherent spin configurations at low temperatures. The correlated paramagnetic phase is also accompanied by specific electrical resistivity anomalies. The *T*-*x* phase diagram determined for the $UCo_{1-x}Ir_xGe$ compounds contrasts with the behavior of the related $URh_{1-x}Ir_xGe$ system, which was reported to exhibit an extended concentration range of stable ferromagnetism in Rh rich compounds and a discontinuous transformation between the ferromagnetic and antiferromagnetic phases at a critical Rh-Ir concentration. The striking difference is tentatively attributed to a) the instability of tiny U moment in the weak itinerant ferromagnet UCoGe induced by substitutional disorder at already very low Ir doping, b) the rather stable U moment in URhGe formed by much less delocalized 5*f* -electrons assisted by weakly varying lattice parameters of $URh_{1-x}Ir_xGe$ compounds.

*Corresponding author: Dávid Hovančík
E-mail: hovancik@mag.mff.cuni.cz

## INTRODUCTION

Orthorhombic U*TX* (*T* - transition metal, *X* – a *p*-electron element) uranium compounds of TiNiSi-type structure are of constant interest due to the extraordinary richness of various magnetic and superconducting states and other strongly correlated electron phenomena. Special attention was paid to URhGe [1] and UCoGe [2] due to the unique coexistence of ferromagnetism (FM) and superconductivity (SC). The evolution of magnetism with increasing distance of U nearest neighbors within the series of UTGe compounds seems to follow Hill's scenario [3-5] when starting by the paramagnetic (PM) ground state of URuGe followed up by ferromagnets UCoGe[2] and URhGe[6] and antiferromagnets (AFM) UIrGe[7, 8], UNiGe[9-11] and UPdGe[12, 13]. Despite this empirical finding, the evolution of magnetism in pseudo-ternary alloy systems is often unexpectedly different. This confirms the fact that the microscopic origin of the magnetism in these compounds is more complex than determined only by the overlaps of the 5*f*-orbitals of the nearest-neighbor U ions. The strong role of competing FM and AFM interactions mediated by the 5*f*-ligand hybridization determining the magnetic ground state was predicted for these compounds [14, 15] as well as for the recently discovered nonmagnetic heavy-fermion superconductor $UTe_2$ [16, 17].

Strong magnetocrystalline anisotropy, which reflects the capture of 5*f*-electron magnetic moments in certain crystallographic directions, appears to be a generic property of uranium magnets. The strong interaction of spatially extended U 5*f* -orbitals with the orbitals of surrounding ligands and the involvement of 5*f*-electrons in bonding[18, 19] imply a mechanism of magnetocrystalline anisotropy based on a two-ion (U-U) interaction which has been theoretically described by B.R. Cooper et al.[20, 21]. The anisotropy of the bonding and 5*f*-ligand hybridization assisted by the strong spin-orbit interaction are the key ingredients. The systematic occurrence of particular types of anisotropy related to the layout of the U ions in crystal lattices in which the U-U coordination is clearly defined suggests an empirical rule that the easy magnetization direction is in the plane perpendicular to the nearest U-U links[11, 22]. It is well documented e.g. by the uniaxial anisotropy with the easy magnetization direction along the *c*-axis in hexagonal UTX compounds with the ZrNiAl-type structure in which the nearest neighbor U atoms are concentrated in basal-plane U sheets[11, 22].

The nearest U atoms in the orthorhombic structure of the TiNiSi-type common to UTGe compounds form zigzag chains running along the *a*-axis, which appears to be the hardest direction of magnetization. However, the situation is not as straightforward as in the hexagonal compounds. The actual type of anisotropy seems to be related to the type of magnetic ground state. The ferromagnets adopt a uniaxial anisotropy[23-25] whilst an orthorhombic anisotropy[26] is characteristic for antiferromagnets[8, 27, 28].

UCoGe is known as an Ising-like weak itinerant ferromagnet with a very reduced ordered moment $\mu_0$ = 0.07 $\mu_B$ oriented along the orthorhombic *c*-axis. The *a*- and *b*-axis linear *M(B)* dependences, respectively, represent very weak paramagnetic signals in directions perpendicular to *c*[29].

UIrGe is an antiferromagnet exhibiting the orthorhombic anisotropy [7,22]. The easy magnetization direction in both the paramagnetic and the AFM state is parallel to the *c*-axis. The magnetization observed in the *b*-axis direction is intermediate between the *c*- and *a*-axis signals. The superconducting phenomena in UCoGe and URhGe seem to be related not only to the ferromagnetic (FM) ground state but also to various paramagnetic modes, which develop at much higher temperatures. However, the nature of these relationships remains unexplained. The unusual features of the FM and paramagnetic states in the uranium FM superconductors have been revealed by studies of their pseudo-ternary alloy systems [1, 30, 31]. UIrGe is isostructural and isoelectronic with UCoGe and URhGe but on contrary, it exhibits an antiferromagnetic (AFM) ground state and doesn't show superconductivity.

It is, of course, of natural interest to see the effect of alloying UIrGe with the two FM superconductors. The $URh_{1-x}Ir_xGe$ case has been studied and results published earlier[15]. The key feature reported for this system is the discontinuity in all the magnetic parameters between the FM and AFM phase typical for the first-order transition induced by the change of Rh/Ir composition ratio at $x_{crit}$

= 0.56.

These aspects motivated us to investigate also the $UCo_{1-x}Ir_xGe$ system by measurements of magnetization, specific heat, and electrical resistivity at various temperatures and magnetic fields. Results of our present study indicate that contrary to the behavior of the $URh_{1-x}Ir_xGe$ system, the $UIr_{1-x}Co_xGe$ compounds have a paramagnetic ground state in a rather wide range of intermediate concentrations between the FM and AFM phase spaces. The specific-heat and magnetization anomalies showing up at elevated temperatures can be understood as effects of strong AFM correlations and/or freezing of incoherent configuration of magnetic moments with possible short-range magnetic order.

Technical issues of this paper (Experimental), results of chemical and crystallographic analysis (Concentrations and lattice parameters), and some supplementary results in graphical and tabular form have been moved to Supplementary Materials and referenced as ([32]).

## RESULTS AND DISCUSSION

### Magnetization

The low-field (0.01 T) thermomagnetic curves of $UCo_{1-x}Ir_xGe$ compounds: in polycrystalline form $x \leq 0.3$ (Fig. 1a, b) monotonously increase at a gradually increasing negative slope with decreasing temperature. The slope of the *M(T)* curves at a corresponding temperature decrease with increasing *x*. The *M(T)* curves obtained on single crystals for $x \leq 0.07$ in the magnetic field applied parallel to the *c*-axis show analogous evolution (Fig. 1a). When increasing Ir concentration for $x \geq 0.4$ in polycrystals (Fig. 1b) and $x \geq 0.24$ in single crystals (Fig. 1c), the *M(T)* curves show a peak. A peak on an *M(T)* curve may appear due to several reasons: a magnetic phase transition to an AFM state, freezing of glassy spins configurations, short-range AFM ordering, or AFM correlations in a paramagnetic phase. The *M(T)* curves have for $x \leq 0.8$ a more or less symmetric bump-shape. For $x > 0.8$ the anomaly is abruptly terminated by an edge and drop on the low-temperature side (Fig. 1b, c) similar to the *M(T)* dependence exhibited by the UIrGe single crystal (also shown in Fig. 1c).

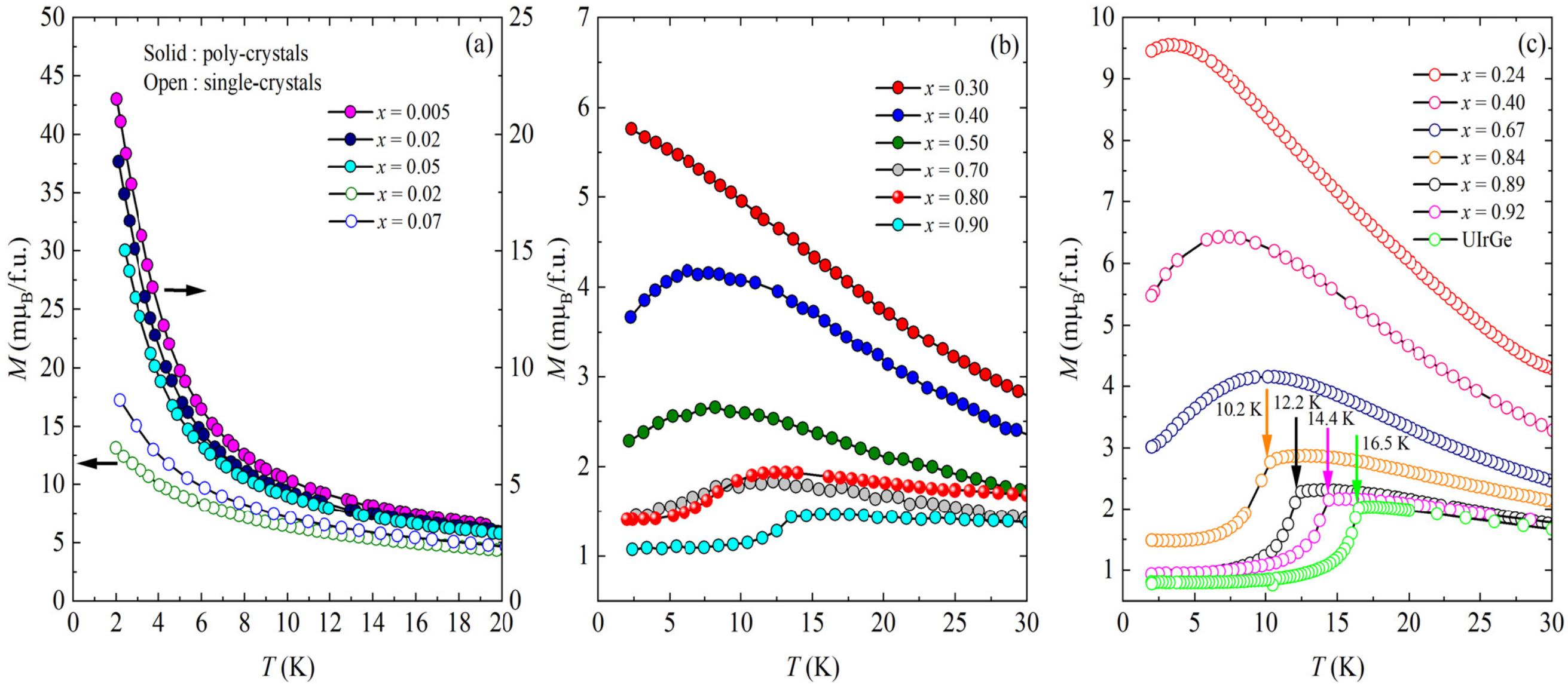

FIG. 1. The thermomagnetic curves measured on selected $UCo_{1-x}Ir_xGe$ samples in $\mu_0 H$ = 100 mT for a) polycrystals with $x \leq 0.05$ and single crystals with $x \leq 0.07$, b) polycrystals with $x \geq 0.3$, and c) single crystals with $x \geq 0.24$ in magnetic fields applied along the c-axis. The vertical arrows and corresponding

labels indicate Néel temperatures determined from specific-heat data measured on (see below) AFM single crystals (for $x \geq 0.84$) in panel (c) and AFM polycrystal ($x = 0.9$) in panel (b).

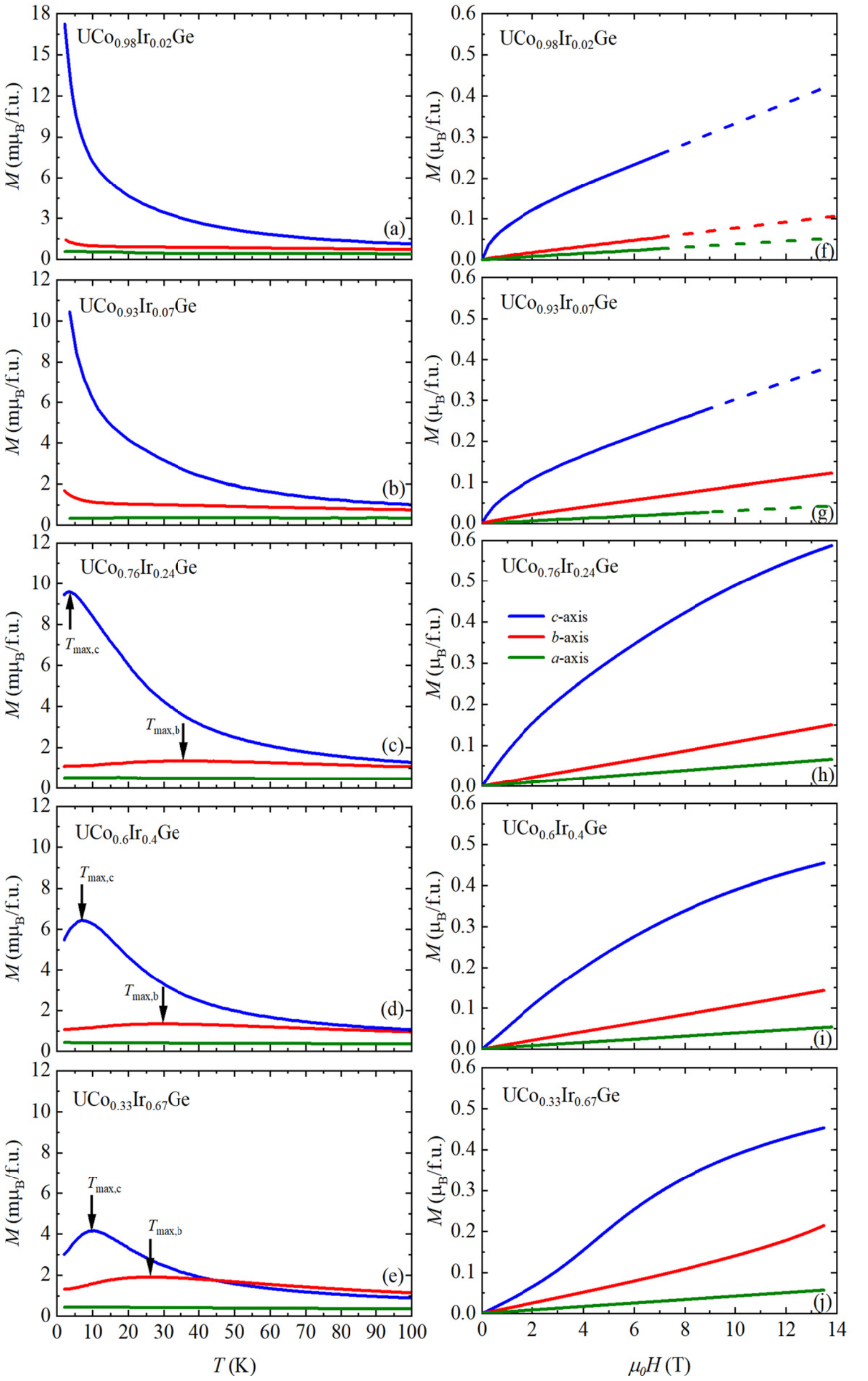

FIG. 2. Left column: The thermomagnetic curves measured on the single-crystals (a) $UCo_{0.98}Ir_{0.02}Ge$, (b) $UCo_{0.98}Ir_{0.02}Ge$, (c) $UCo_{0.76}Ir_{0.24}Ge$, (d) $UCo_{0.6}Ir_{0.4}Ge$ (e) $UCo_{0.23}Ir_{0.67}Ge$ in a magnetic field of 100 mT applied parallel to the *a*-, *b*- and *c*-axis, respectively. The temperatures of *M(T)* maxima are marked by vertical arrows and corresponding labels. Right column: the corresponding magnetization curves measured at 2 K in magnetic fields applied parallel to the *a*-, *b*- and *c*-axis, respectively. The dashed lines represent the linear extrapolation from 7 (9) to 14 T.

The above features of the *M(T)* curves are naturally more clearly seen in the results measured on single-crystals (see Fig. 1c) in the magnetic field applied parallel to the easy magnetization direction. The $UCo_{0.98}Ir_{0.02}Ge$ crystal represents a compound that remains paramagnetic down to the lowest measured temperature (400 mK) but appears in the vicinity of the onset of ferromagnetism with increasing Co concentration. Then follow the paramagnetic compounds in the intermediate concentration range ($x$ = 0.24, …..0.67) exhibiting a broad peak on the *M(T)* curve mentioned above for polycrystals. The crystals with Ir concentrations $x \geq 0.84$ behave qualitatively similar to the antiferromagnet UIrGe (x = 1.0).

Data measured on individual single crystals (shown in detail in Figs. 2 and 3) enable us to see that the magnetization of $UCo_{1-x}Ir_xGe$ compounds is strongly anisotropic and how the anisotropy develops throughout the pseudoternary system. The *M(B)* data measured on the $UCo_{1-x}Ir_xGe$ single crystals at 2K reveal that the *a*-axis is in all studied compounds the hard magnetization direction characterized by the weak linear *M(B)* increase reaching ≈ 0.05 $\mu_B$/f.u. in an external magnetic field of 14 T irrespective of the magnetic ground state. In the sequence of crystals with increasing *x*, it can be seen that the crystals with $x \leq 0.24$ exhibit uniaxial anisotropy comparable to that observed in the uniaxial ferromagnet UCoGe with the *c*-axis easy direction of magnetization[33]. It is characterized by the relation of the 2-K magnetization measured parallel to the *c*-, *b*- and *a*-axis, respectively, $M^c >> M^b \approx 2M^a$; both $M^a(B)$ and $M^b(B)$ are linear functions. With increasing *x* beyond 0.24 the ratio $M^b(B) : M^a(B)$ gradually increases (see also Table S2[32]). This indicates the change of the anisotropy from uniaxial to orthorhombic. For $x \geq 0.67$ $M^b(B)$ becomes nonlinear. For $x$ = 0.84 and 0.89 we can see that both the *b*- and *c*-axis are easy magnetization directions, i.e. the *b*-*c* plane is the easy plane. For $x$ = 0.92 and 1.0 (UIrGe) the *b*-axis is the easy magnetization direction. This evolution of anisotropy is reflected also in the thermomagnetic curves measured in fields applied along the three main crystallographic axes (see. Figs 3 and 4).

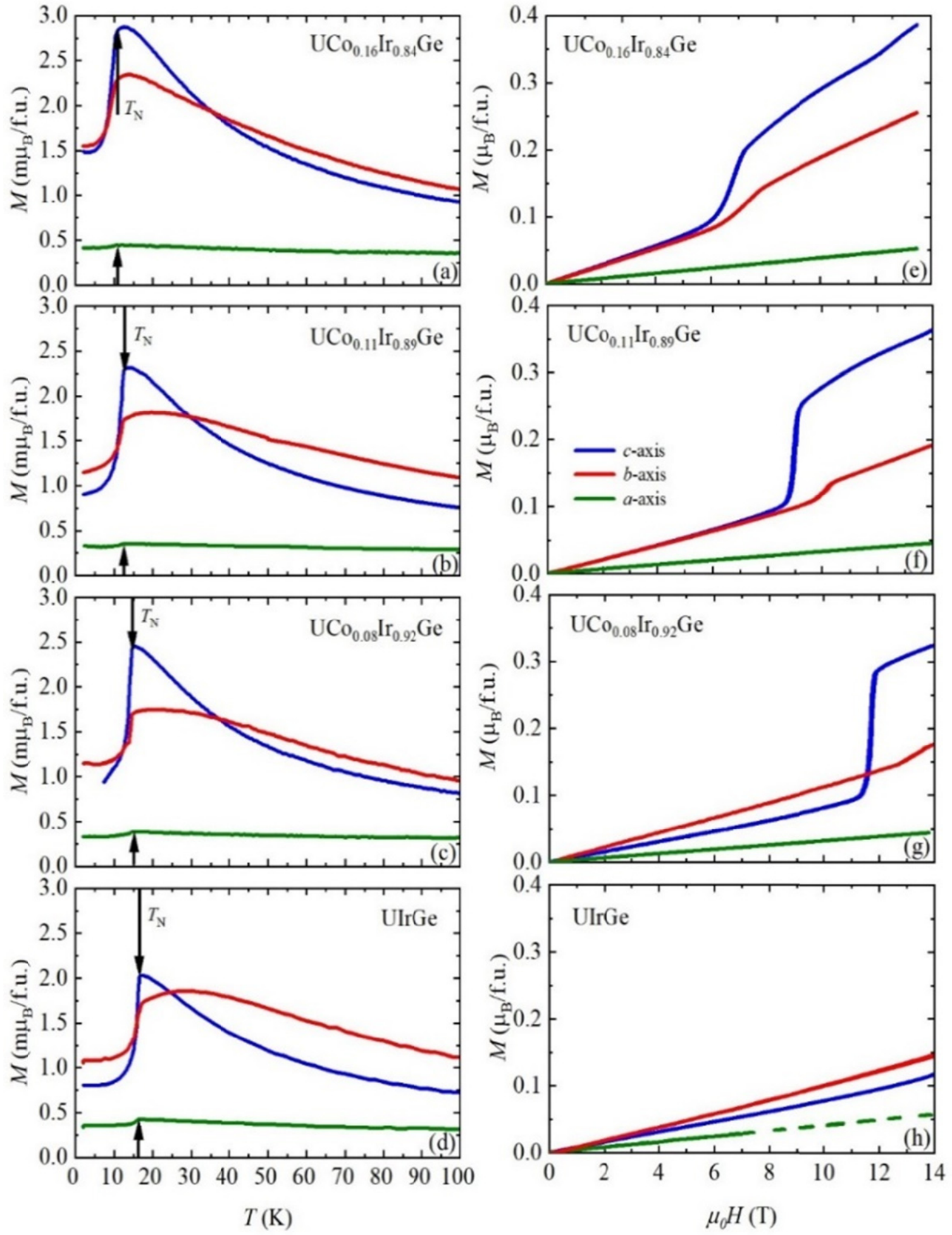

FIG. 3. Left column: the thermomagnetic curves measured on the single-crystals (a) $UCo_{0.16}Ir_{0.84}Ge$, (b) $UCo_{0.11}Ir_{0.89}Ge$, (c) $UCo_{0.078}Ir_{0.92}Ge$, and (d) UIrGe in a magnetic field of 100 mT applied parallel to the *a*-,*b*- and *c*-axis, respectively. The Néel temperatures determined from specific-heat data are marked by vertical arrows and corresponding values Right column: the corresponding magnetization curves measured at 2 K in magnetic fields applied parallel to the *a*-, *b*- and *c*-axis, respectively. The dashed line is the linear extrapolation from 7 to 14 T.

The low-field concave curvature of the 2-K magnetization isotherms (Fig. S2a[32] and Fig. 2) measured for $x \leq 0.05$ on polycrystals and for $x = 0.02$ and 0.07 on single crystals (the *c*-axis) followed by a nearly linear increase in higher fields signals the proximity of itinerant ferromagnetism observed in UCoGe. The absence of ferromagnetism at temperatures down to 2 K in these samples has been confirmed by Arrott plots which are shown in Figs S3 and S4[32]. To find a possible FM transition at temperatures below, we have measured the AC susceptibility of the low-*x* samples using a provisional setup consisting of driving and pick-up coils wound around the measured sample and attached to the $^3$He stick of PPMS enabling measurements at temperatures down to 400 mK. The measurements revealed a peak in the temperature dependence of both, the real ($\chi$') and imaginary ($\chi$'') component of AC susceptibility at ≈ 1 K (see Fig. S5[32]) for samples with $x = 0.005$, 0.01 that can be understood as the onset of ferromagnetism at this temperature. No signs of bulk superconductivity were observed in the AC susceptibility data measured on samples with $x < 0.01$. On samples with $x > 0.01$ no anomaly was observed which would indicate an FM or superconducting transition at temperatures down to 400 mK, neither in $\chi'(T)$ nor $\chi''(T)$ data.

The low-field part of 2-K isotherms (see Fig. S2b for polycrystals[32] and Fig. 3 for single crystals in *B*//*c*-axis) gradually straightens with increasing Ir content up to $x = 0.6$ due to gradually enhanced involvement of AFM interactions. The increasing involvement of AFM interaction is reflected in the slight S-shape of *M(B)* curve for the polycrystalline sample with $x = 0.7$ (Fig. S2(b)[32]) and the single crystal with $x = 0.67$ for *B*//*c*-axis.

The 2-K magnetization isotherms measured on the single crystals with $x = 0.24$, 0.4 and 0.67 (Figs.2(c) 3(c, d, e)) and polycrystals with $x = 0.4$, 0.5, 0.7 and 0.8 (Fig. 2) are, in our opinion, characteristic for a strongly anisotropic paramagnet. Thermomagnetic curves (Fig. 2) show a broad peak (for single crystals only the curves for *B*//*b* and *B*//*c*) characteristic of a considerable involvement of AFM correlations. The $M^c(T)$ values dominate whereas the $M^b(T)$ peak is much broader and located at higher temperatures.

Further increasing Ir concentration above $x = 0.8$ (see Fig. 1) lead to antiferromagnetism. The AFM transition is reflected as a sudden decrease of magnetization with decreasing temperature in the $M^c(T)$ and $M^b(T)$ dependences seen in Figs. 1, 3. The AFM ordering is documented also by the metamagnetic transitions observed in the 2-K $M^b(B)$ and $M^c(B)$ isotherms.

The temperature dependences of paramagnetic susceptibility calculated from $M^c(T)$ and $M^b(T)$ dependences follow at temperatures $T > 50$ K the modified Curie-Weiss law

$$\chi = \frac{N_A \mu_{eff}^2}{3k_B(T-\theta_p)} + \chi_0 \qquad (1)$$

with parameters ($\mu_{\rm eff}$ – effective magnetic moment, $\theta_{\rm p}$ – paramagnetic Curie temperature, $\chi_0$ – temperature-independent susceptibility) shown in Table S3[32]. $N_{\rm A}$ = Avogadro number, $k_{\rm B}$ = Boltzman's constant. We are well aware that the values of the corresponding parameters cannot be considered to correctly describe the paramagnetic state in such complex compounds as $UCo_{1-x}Ir_xGe$. However, a closer examination of Table S3[32] reveals some tendencies that can be considered in the discussion in conjunction with other experimental results. From fitting the easy-magnetization-direction (*c*-axis) paramagnetic susceptibility we can see (a) that the $\mu_{\rm eff}$ values appear in a relatively narrow corridor around the mean value of 1.83 $\mu_B$/f.u., (b) The small positive $\theta_p^c$ values show the maximum at $x = 0.24$ and decrease with increasing the Ir concentration towards $\theta_p^c \approx 0$ K at $x \leq 0.67$. Then $\theta_p^c$–value drops to -11 K at $x = 0.84$ and remains practically constant with increasing Ir concentration up to 1.0 (UIrGe). The concentration interval characterized by negative $\theta_p^c$–values coincides with the interval of stability of antiferromagnetism in $UCo_{1-x}Ir_xGe$ compounds. The values of the $b$-axis effective magnetic moment are around the mean value of 2.52 $\mu_B$/f.u., which is almost equal to the expected value for the free $U^{3+}$ ion (2.54 $\mu_B$). The very large negative $\theta_p^b$-value of -164 K for $x$ = 0.02 reflects nearly temperature-independent $b$-axis susceptibility due to the strong uniaxial anisotropy with the easy-axis direction perpendicular to the $a$-$b$ plane. When increasing the Ir concentration further the $\theta_p^b$-values become gradually reduced down to - 39 K at $x$ = 0.67 reflecting the transformation of anisotropy from uniaxial to orthorhombic. For $x \geq 0.84$ where we find antiferromagnets, and the negative $\theta_p^b$-values settle in a corridor - 35 ± 7 K.

The $a$-axis susceptibility is very slightly increasing with decreasing temperature in all studied crystals. It drops by about 10 % at the AFM transition with cooling. This anomaly is small nevertheless ubiquitous in all AFM crystals and indicates that the $a$-axis component plays a non-negligible role in non-collinear magnetic structures of the AFM $UCo_{1-x}Ir_xGe$ compounds similar to the isostructural case of UNiGe[34-36].

It is worth noting that magnetization data measured on polycrystals of materials with considerable magnetocrystalline anisotropy, such as the $UCo_{1-x}Ir_xGe$ compounds, provide only limited informative value. We measure an average value of actual magnetizations of randomly oriented small single-crystalline grains projected on the axis of the magnetization detection system. If we measure the magnetization of a polycrystal in the direction of applied magnetic field sufficiently smaller than the anisotropy field the measured values represent a spatial average of the easy-axis magnetization ($M^{easy}$). In the case of materials with strong uniaxial anisotropy $M^{polycr} = 0.5\ M^{easy}$. In our case, we consider the temperature of $M^{polycr}(T)$ as the temperature of $M^c(T)$ maximum.

**Specific heat**

The specific-heat data measured for all available polycrystalline and single-crystalline samples are shown in Fig. 4. In this context, it is worth noting that the specific heat is an isotropic property of material if measured in the absence of a magnetic field. The specific-heat behavior of the material of the same composition can then be considered the same regardless of whether they are measured on a polycrystal or a single crystal. We take this fact into account in the following discussion.

The $T_C$-related anomaly and the steep increase below 1 K associated with the onset of superconductivity in the $C_P/T$ vs. $T$ data observed for UCoGe are wiped out already from data measured on the sample with $x$ = 0.005). The convex $C_P/T$ vs. $T$ curves for $x \geq 0.05$ gradually straighten with increasing $x$ up to 0.5 where a broad bump appears on an almost straight $C_P/T$ vs. $T$ dependence. The bump develops gradually with further increasing Ir content. For $x$ = 0.84 a λ peak reflecting a second-order magnetic phase transition at 10.2 K appears. The peak qualitatively similar to the $T_N$-related anomaly observed for UIrGe gradually grows and moves to higher temperatures with increasing Ir concentration.

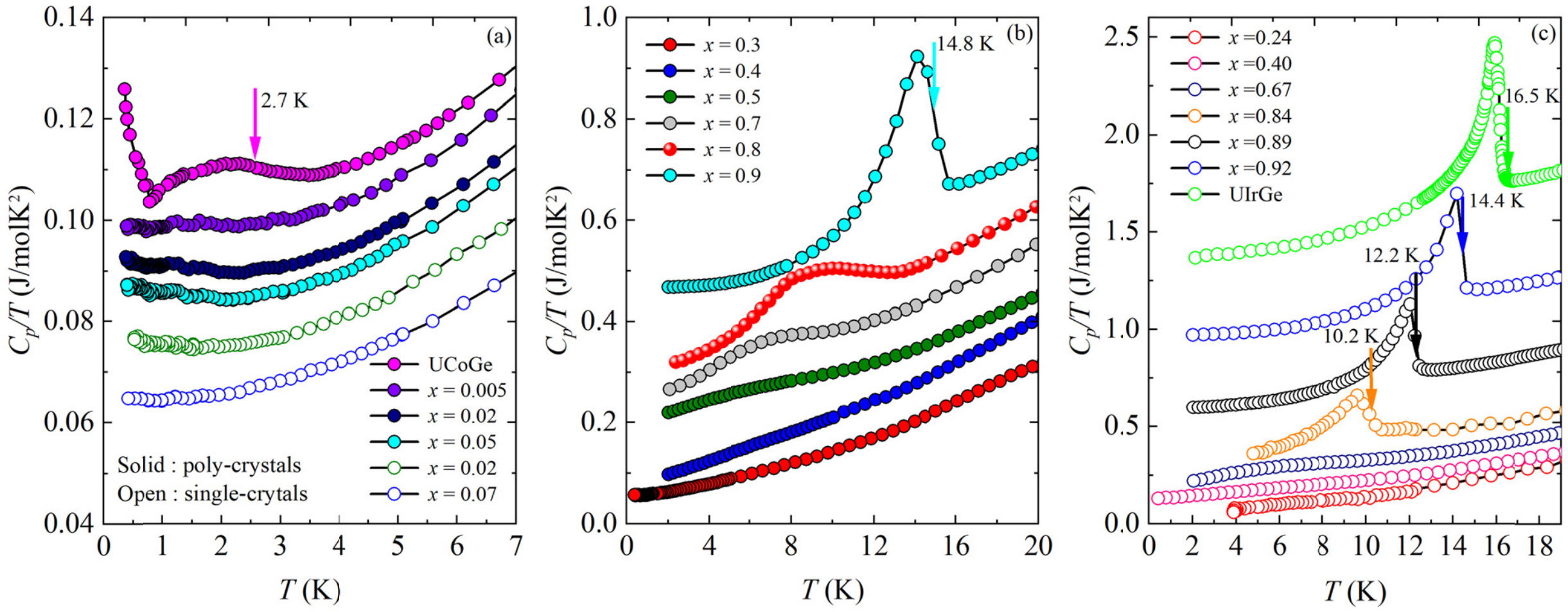

FIG. 4. The temperature dependences of specific heat ($C_P/T$ vs $T$ plots) in zero external magnetic field of $UCo_{1-x}Ir_xGe$ compounds for a) polycrystals with $x \leq 0.05$ and single crystals with $x \leq 0.07$, b) polycrystals with $x \geq 0.3$ and c) single crystals with $x \geq 0.24$ in magnetic fields applied along the c-axis. The vertical arrows and corresponding labels indicate Néel temperatures of AFM single crystals (for $x \geq 0.84$) in panel (c) and AFM polycrystal ($x$ = 0.9) in panel (b). For the sake of clarity, the curves are gradually shifted for a) by a linear offset($x$) decreasing with increasing $x$, and for b) and c) by an offset($x$) increasing with increasing $x$. Data for UCoGe are taken from Ref. [19].

To determine the magnetic entropy $S_{mag}$ we have subtracted the phonon $C_{ph}$ and electron $C_{el}$ contribution from the experimental $C_P$ data as shown in Fig. 7(a). Since the simple Debye ~ $T^2$ dependence does not provide a usable fit a general polynomial function with the dominant quadratic term was used. Fig. 7(a) shows a representative example. It is seen in Fig. 7(b) that $S_{mag}$ rapidly decreases with decreasing $x$ from 0.19 $R$ln2 for $x$ = 1.0 (UIrGe) down to ~ 0.07 $R$ln2 for $x$ = 0.8, which most likely reflects rapid suppression of the AFM order with Co substitution for Ir. For $x$ = 0.8, the anomaly in the $C_P/T$ vs $T$ dependence has a form of a broad bump which is most probably not connected with an AFM phase transition but it reflects critical AFM fluctuations, short-range AFM ordering, or frozen spin-glass-like state. This picture is corroborated by observing the magnetization (broad $M(T)$ peak in Fig. 1a and slight $M(B)$ S-shape in Fig. S2b[32]) and electrical resistivity (rapid $\rho(T)$ increase with decreasing temperature below 10 K) behavior. The further decrease of $S_{mag}(x)$ with decreasing $x$ from 0.7 to 0.4 is much slower and $S_{mag}$ vanishes around 0.3.

The overall evolution of $S_{mag}$ with decreasing Ir content is in line with the evolution of magnetization and specific-heat anomalies and may be tentatively understood in terms of suppression of long-range AFM order at a critical concentration $x_c$ near to 0.8. The tiny broad bumps in $C_P/T$ vs $T$ observed for 0.3 < $x$ ≤ 0.7 may be understood in terms of short-range magnetic ordering, freezing of a glassy configuration of spins, or AFM correlations in the PM state similar to e.g. the $CeRu_2Si_2$ case [37].

The large difference between $S_{mag}$ of UCoGe and UIrGe in conjunction with a qualitatively similar difference of U magnetic moments in the two compounds highlights the fact that the 5$f$-electron states in UCoGe are considerably more delocalized than in UIrGe.

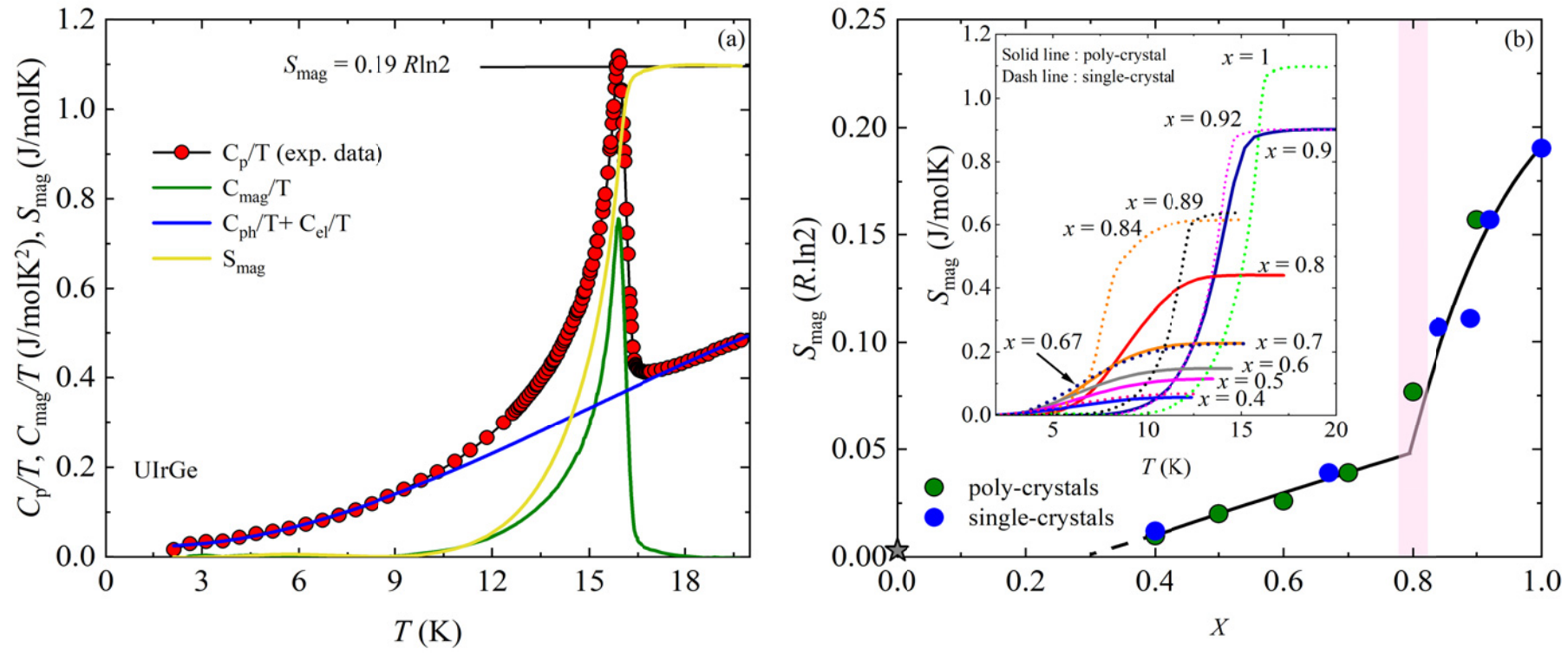

FIG. 5. (a) Temperature dependence of specific heat and its components ($C_P/T$ vs $T$ plots) and magnetic entropy $S_{mag}$ of UIrGe and (b) the value of $S_{mag}$ as a function of $x$ expressed in units $R$ln2 (polycrystals = black points, single crystals = red points). The inset shows $S_{mag}$ versus $T$ for AFM compounds. The value of $S_{mag}$ for UCoGe (marked by the star) was taken from Ref.[38, 39].

**Electrical resistivity**

The temperature dependencies of electrical resistivity, $\rho(T)$, measured on polycrystalline samples are shown in Fig. S6 [32] in $\rho(T)$ plots. No anomaly which could be attributed to a magnetic phase transition to FM or AFM state can be identified in the displayed $\rho(T)$, curves except for the sample with $x = 0.9$, where the dramatic drop with decreasing temperature is observed below 25 K ($T_N$ = 14.8 K has been determined from specific heat data shown in Fig. 4b). A shallow minimum at around 7 K is observed similar to the minimum observed in pure UIrGe [38, 40]. The reverse behavior of the sample with $x = 0.8$ exhibiting rapidly increasing resistance with the decreasing temperature that was mentioned already above is probably reflecting critical AFM fluctuations, short-range AFM ordering, or a frozen spin-glass-like state. Similar, but less pronounced $\rho(T)$ behavior is seen for $x$ = 0.7. When further decreasing Ir concentration a broad peak at 25 K in the $\rho(T)$ dependence appears for $x$ = 0.6. A low and broad $\rho(T)$ peak has been found also at temperatures in the interval 26 – 33 K (see Table S4) for the intermediate compositions $x$ = 0.1 - 0.4.

The dramatic drop of electrical resistivity observed only for samples with $x$ = 0.005 and 0.01 when cooling below 0.5 K (Fig. S7[32]) can be understood as the onset of the superconducting transition. Nevertheless, a zero resistivity was not reached down to 0.4 K. Bulk superconductivity in these samples can be expected below 0.4 K.

The low-temperature $\rho(T)$ data were fitted with the expression:

$$\rho = \rho(0) + AT^n \qquad (2),$$

where $\rho(0)$ is the residual resistivity. The values of the exponent $n$ listed in Table S5[32] are in most cases much lower than 2, which is usually considered as a hallmark of non-Fermi liquid behavior.

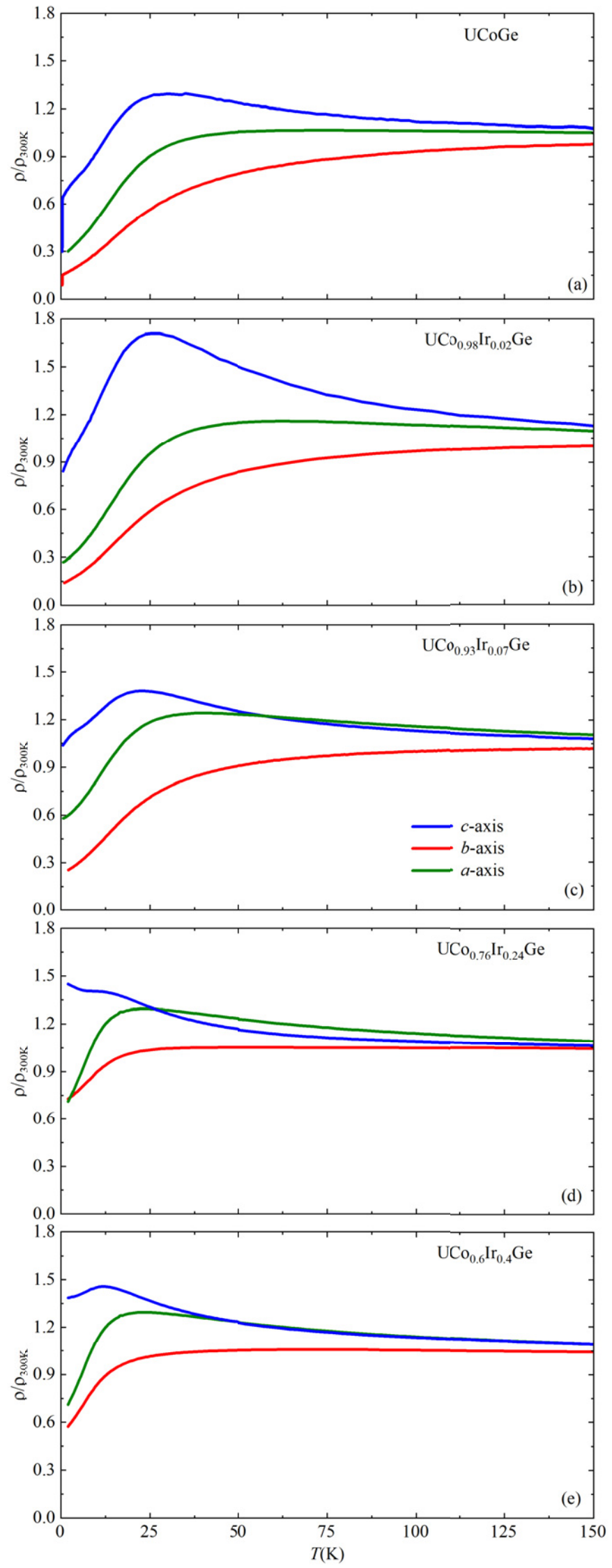

FIG. 6. Temperature dependence of relative electrical resistivity on the single-crystals (a) UCoGe, (b) $UCo_{0.98}Ir_{0.02}Ge$, (c) $UCo_{0.93}Ir_{0.07}Ge$, (d) $UCo_{0.76}Ir_{0.24}Ge$ and (e) $UCo_{0.6}Ir_{0.4}Ge$ for electrical current parallel to the *a*- (blue), *b*- (red) and *c*-axis (green), respectively.

The $\rho(T)$ dependences measured on single crystals for current parallel to each main crystallographic axis ($\rho^a(T)$, $\rho^b(T)$, $\rho^c(T)$, for *i*//*a*, *i*//*b*, *i*//*c*, respectively) are displayed in Figs. 6. and 7. These data document the anisotropy of temperature dependence of resistivityof the studied compounds. The $\rho(T)$ data measured on polycrystals are the result of multiple processes of intragrain and intergrain transport by a conglomerate of quasi-randomly oriented grains between stress contacts on the measured sample. It is clear that resistivity data measured on polycrystals of materials with anisotropic resistivity, such as $UCo_{1-x}Ir_xGe$ compounds, do not have much informative value regarding intrinsic electrical transport behavior.

This fact can be documented by comparing the $\rho(T)$ dependences measured on a polycrystal and corresponding $\rho^a(T)$, $\rho^b(T)$, $\rho^c(T)$ dependences measured on a single crystal of the corresponding composition, e.g. $x = 0.02$ or 0.4.

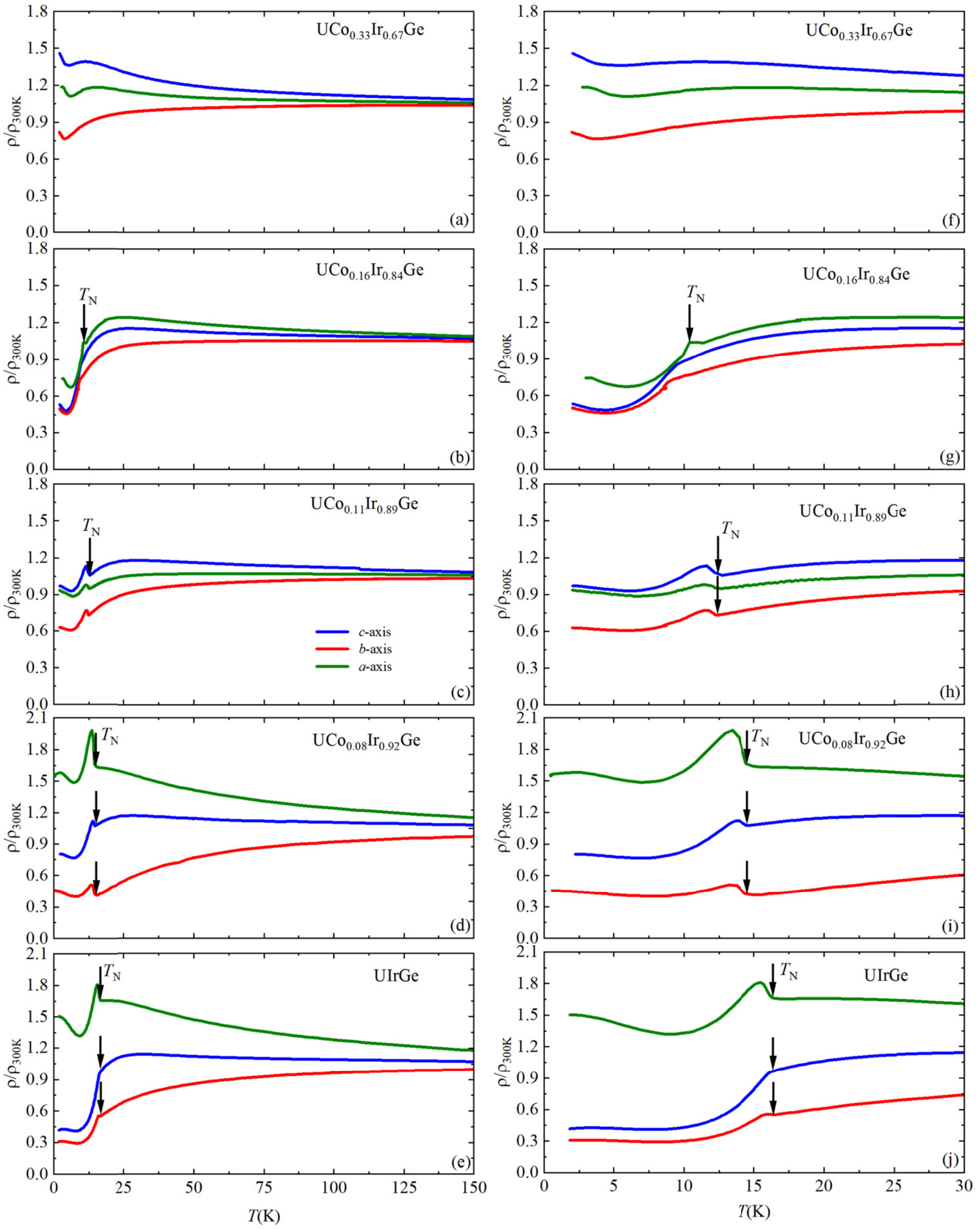

FIG. 7.: Left column: Temperature dependence of relative electrical resistivity on the single-crystals (a) $UCo_{0.23}Ir_{0.67}Ge$ (b) $UCo_{0.16}Ir_{0.84}Ge$, (c) $UCo_{0.11}Ir_{0.89}Ge$ (d) $UCo_{0.078}Ir_{0.92}Ge$ and (e) UIrGe for electrical current parallel to the *a*- (blue), *b*- (red) and *c*-axis (green), respectively. Right column: Low temperature details of the corresponding $\rho^a(T)$, $\rho^b(T)$, $\rho^c(T)$ plots, respectively, displayed in the left column.

The resistivity of the $x = 0.02$ polycrystal remains constant with cooling from 100 to 50 K and then decreases with further similarly decreasing temperature as the $\rho^a(T)$ dependence. But the $\rho^c(T)$ dependence measured on a single crystal exhibits a pronounced broad peak at 25 K which reflects scattering of conduction electrons on anisotropic magnetic fluctuations along the *c*-axis. A similar but considerably lower $\rho^c(T)$ anomaly is observed for UCoGe and $UCo_{0.93}Ir_{0.07}Ge$. This means that the scattering of conduction electrons on magnetic fluctuations is the strongest in $UCo_{0.98}Ir_{0.02}Ge$. This result in conjunction with the practically linear $\rho^c(T)$ dependence at lowest temperatures can be most probably associated with the proximity of the onset of ferromagnetism which is expected at a somewhat lower Ir concentration than $x = 0.02$.

The $\rho^c(T)$ dependences measured on single crystals with $x = 0.24$ and 0.4 are also anomalous reflecting some role of scattering of conduction electrons on the magnetic fluctuation at low temperatures and also the knees on the $\rho^a(T)$ and $\rho^b(T)$ becoming more pronounced are shifted to lower temperatures which can be due to growing involvement of AFM interactions in the hierarchy of exchange interactions. The low-temperature upturns of $\rho^a(T)$, $\rho^b(T)$, and $\rho^c(T)$ curves, as well as the S-shape of the corresponding $M^c(H)$ curve in Fig. 2, indicate strong involvement of AFM correlations in these phenomena.

Finally the of $\rho^a(T)$, $\rho^b(T)$, and $\rho^c(T)$ data obtained on single crystals with $x \geq 0.84$ are typical for an antiferromagnet with anomalies similar to those observed at $T_N$ on AFM rare-earth metals with spiral spin structures[41] but also on the famous itinerant electron antiferromagnet Cr[42, 43]. In the first case, the scenario considering that the spiral spin structures cause an exchange field at the conduction electrons with a lower symmetry than that of the crystal lattice is applied. This introduces new boundaries in the Brillouin zone and distorts the Fermi surface. This distortion and the scattering of the conduction electrons by the spin disorder give rise to the specific anomaly at $T_N$. This approach may probably be considered also on non-collinear antiferromagnets as UIrGe and its Co-doped variants. Considering the theory presented by Elliot and Wedgwood[41] we can expect that the particular magnetic structure has a different periodicity than the crystal structure in the direction in which the applied electric current is accompanied by this specific $\rho(T)$ anomaly at $T_N$. However, this statement conflicts with the proposed AFM structure in UIrGe[44, 45], which has been proposed non-collinear but commensurate with the crystallographic unit cell. This is another argument for re-investigation of the magnetic structure of UIrGe by neutron diffraction.

A closer inspection of $\rho(T)$ dependences connected with cooling in the AFM state is the upturn at temperatures below 10 K. The size of this effect for fixed current direction has been found sample dependent. When applying a magnetic field larger than the critical field of metamagnetic transition by which the AFM structure is destroyed the upturn is suppressed and the resistivity with decreasing temperature below 10 K continuously decreases. Considering these findings we tentatively attribute this $\rho(T)$ upturn to some defects (e.g. stacking faults, spin slips reported e.g. in the case of CePtSn[46, 47]) of the AFM structure which can be caused by some crystal structure defects.

## Magnetic phase diagram

The magnetization, specific heat, and resistivity data presented above allow us to sketch a tentative *T-x* magnetic-phase diagram for the $UCo_{1-x}Ir_xGe$ system shown in Fig. 8. Part of the information on the parent UCoGe and UIrGe compounds have been taken from the relevant literature sources [35, 22].

The AC susceptibility and electrical resistivity measurements down to 0.4 K revealed that $T_C$ of the weak itinerant 5*f*-electron ferromagnetism is rapidly reduced already with slight Ir substitution for Co (to ≈ 1 K for $x = 0.01$) and the superconductivity persist only up to $x = 0.01$. At this Ir concentration, the superconductivity seems to also persist as documented by the drop of electrical resistivity with cooling below 0.5 K but $T_s$ is decreased well below 0.4 K. No signs of ferromagnetism and antiferromagnetism can be traced by resistivity and AC susceptibility at temperatures down to 0.4 K. On the other hand, the AFM ordering in UIrGe survives at least up to the substitution of 16% Co for Ir in UIrGe. We can see that $T_N$ steeply decreases with decreasing *x* from 1 to 0.84 as documented by results obtained on single crystals with $x \geq 0.84$. Measurements for $x = 0.8$ could only be performed on polycrystalline samples. Nevertheless, we believe that the measured and behaviors of magnetization, specific heat, and resistivity discussed above are quite convincing to conclude that the compound of this composition does not become magnetically (AFM) ordered i.e the critical composition for antiferromagnetism in the $UCo_{1-x}Ir_xG$ system is $0.8 < x_{crit} < 0.84$. The diagram also shows the critical magnetic fields of the metamagnetic transition leading to the suppression of the AFM order. The concentration dependences of these critical fields (applied along the *c* and *b* axis, respectively) also point to this concentration interval.

The samples $0.8 \geq x \geq 0.24$ are characterized by broad quasi-symmetric peaks in the *M* vs *T* and very broad bumps in $C_P/T$ vs $T$ dependences declining at a much lower rate with decreasing $x$ than the decrease of $T_N$ (see Figs. 6 and 8). One cannot entirely exclude the possibility of specific AFM phases in this concentration range of the $UCo_{1-x}Ir_xGe$ system similar to those proposed in the case of the $UCo_{1-x}Pd_xGe$ compounds [48]. However, we are more inclined to assume that the highest substitution disorder (see also [35]) in the intermediate concentration range causes the distribution of exchange interactions that lead to complex magnetic correlations in the paramagnetic state (correlated paramagnet – (CPM)) and ultimately to the freezing of incoherent spin arrangements (spin-glass-like, cluster glass [49] with possible short-range magnetic order (SRMO)). These phenomena lead to responses of the system like these we have observed in the concentration range $0.8 \geq x \geq 0.24$. One should be aware that this situation is caused also by a mixture of two T-metals with 3*d* and 5*d* valence electrons of significantly different bandwidths and different strengths of spin-orbit interaction. The magnetism in $UCo_{1-x}Ir_xGe$ compounds is further complicated by the strong and complex magnetocrystalline anisotropy that alters with various alloying of the mother compounds.

The paramagnetic state of samples on the Co-rich side ($0.02 \leq x \leq 0.2$) was confirmed both by missing anomalies in specific-heat and magnetic susceptibility data measured down to 0.4 K. The broad peak in the $\rho^c(T)$ dependence measured on $UCo_{1-x}Ir_xGe$ single crystals with $x = 0.02$ and 0.07 at 25 K are analogous to the $\rho^c(T)$ anomaly observed for UCoGe[50], where it is due to the scattering of conduction electrons on anisotropic ferromagnetic fluctuations along the *c*-axis. In this respect, we tentatively attribute the behavior of compounds with $0.02 \leq x \leq 0.2$ to paramagnetism with FM fluctuations. In the case of compounds with $0.24 \leq x \leq 0.8$, we are more inclined to assume paramagnetism influenced by AFM correlations especially due to peaks observed in the $M^c(T)$ and $M^b(T)$ dependences.

The temperatures of $M^c(T)$ and $M^b(T)$ maxima observed for single crystals with $x \geq 0.24$ labeled as $T_{max,c}$, and $T_{max,b}$, respectively, are also displayed in the magnetic phase diagram. In the sense of the aforementioned notes on the relation between the magnetization measured on a polycrystal and a single crystal along the easy magnetization axis, we consider also the temperatures of $M(T)$ maxima measured on polycrystals with $0.3 \leq x \leq 0.8$ as corresponding $T_{max,c}$ values.

The $T_{max}$ values are taken as temperatures that characterize the crossover from CPM to normal PM regime in the paramagnetic phase[51, 52]. The concentration dependence of $T_{max,b}$, which was introduced in Ref. [7] for pure UIrGe has two parts, which are most probably separated by critical concentration for antiferromagnetism which is slightly higher than $x = 0.8$. UIrGe is characterized by $T_{max,b} = 29$ K[7] that is the almost double value of $T_N$ (= 16.5 K). $T_N$ rapidly decreases with decreasing Ir content but $T_{max,b}$ falls at a more than double rate. At $x = 0.84$ $T_{max,b}$ reaches its minimum value and then it increases with decreasing $x$ and approaching the value of 37.5 K observed for UCoGe[53].

In the AFM range the $T_{max,c}$ values are practically equal to $T_N$. For $x$ decreasing from 0.8 the $T_{max,c}$ values slowly decrease to 4 K at $x = 0.24$ with the tendency to vanish in the interval $0.1 \leq x \leq 0.2$.

Experiments that can address the microscopic aspects of magnetism in $UCo_{1-x}Ir_xGe$ compounds directly (magnetic neutron and X-ray scattering, μSR in particular) could provide details concerning the driving mechanism of the specific phenomena in the CPM regime.

The resulting *T*-*x* phase diagram of the $UCo_{1-x}Ir_xGe$ system in Fig. 8 is characterized by a very narrow concentration interval of stability of ferromagnetism and a wide intermediate concentration region without magnetic ordering. This contrasts the evolution of magnetism in the $URh_{1-x}Ir_xGe$ system reported to exhibit an extended range of stable ferromagnetism in Rh rich compounds up to a discontinuous transformation (typical for a first-order transition) between the FM and AFM phases of parent compounds at a critical concentration $x_{crit} = 0.56$ [15].

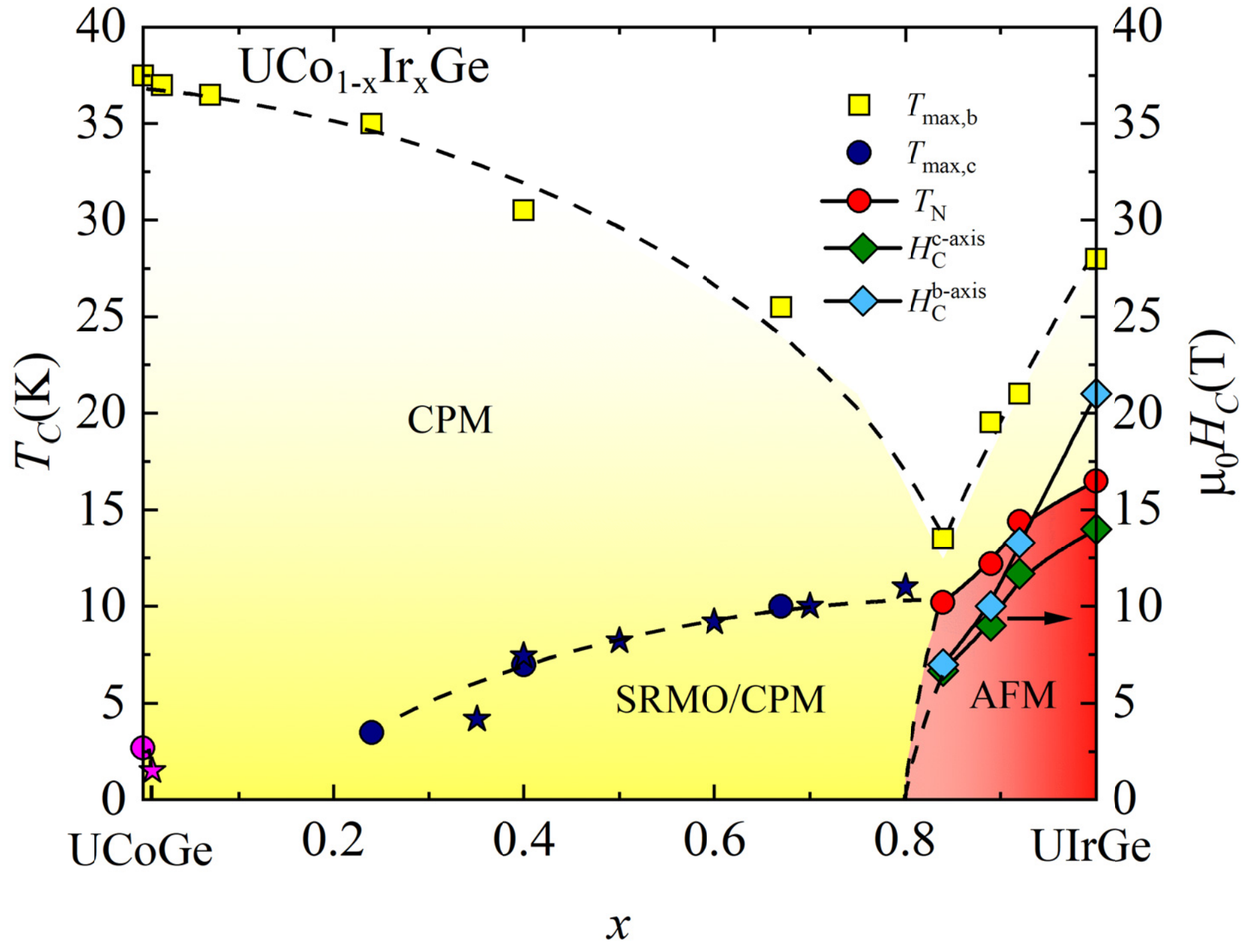

FIG. 8. *T*-*x* magnetic phase diagram of $UCo_{1-x}Ir_xGe$ system. The temperatures marked by magenta/navy stars depicting $T_C/T_{max,c}$ were obtained on polycrystals. The data for UCoGe (marked by a magenta circle) are taken from Ref.[33].

Finding a plausible explanation for the striking difference between the T-x phase diagrams of $UCo_{1-x}Ir_xGe$ and $URh_{1-x}Ir_xGe$ is not an easy task. We can try to approach this problem by discussing the mechanisms affecting the key components of magnetism in uranium intermetallic compounds, uranium magnetic moments, and exchange interactions between them with a special focus on the studied systems.

The 5*f*-electron uranium wave functions, which are propagated in space (unlike 4*f*-electron orbitals deeply buried in the core of lanthanide ions), overlap and interact with the 5*f*-orbitals of neighboring U ions (5*f*-5*f* overlap) and hybridize with valence electron orbitals of nonuranium ligands (5*f*-ligand hybridization)[54]. As a result, the 5*f*-orbitals lose in compounds their atomic character and the U magnetic moments are reduced compared to free U-ion moments ($U^{3+}$ or $U^{4+}$). The large 5*f*-5*f* overlap by rule prevents the formation of a permanent atomic 5*f*-electron magnetic moment in materials in which the distance of nearest-neighbor U atoms (called $d_{U-U}$ ) is smaller than the Hill limit (340-360 pm)[55]. The 5*f*-ligand hybridization has more subtle effects on magnetism which show up in the lower U-content compounds where the ligands surrounding U ions prevent the direct U-U bonds[54]. The direct overlap of U 5*f* -wave functions is also responsible for the direct U-U exchange interaction, while the 5*f*-ligand hybridization mediates the indirect exchange interaction between U ion moments separated by the involved ligand.

The strong spin-orbit interaction in uranium ions induces an orbital magnetic moment that dominates the spin moment. This happens in all so far investigated magnetic U materials even in the cases of weak itinerant ferromagnets like $UNi_2$[56, 57] and UCoGe[58, 59].

The ferromagnetism in UCoGe is characterized by the low $T_C$ value (= 2.7 K) and an extremely reduced spontaneous magnetic moment $\mu_s = 0.07\ \mu_B$ at 2 K [29]. The ferromagnetic ordering in UCoGe is suppressed by the application of low hydrostatic pressure of only ≈ 1 GPa[60-62]. The small magnetic moment and rapid suppression of $T_C$ in hydrostatic pressure are hallmarks of weak itinerant ferromagnetism.

URhGe becomes ferromagnetic at higher $T_C$ (= 9.6 K) and the magnetic moment of $\mu_s \approx 0.4\ \mu_B$ is also reduced but still much larger than that in UCoGe[6, 23]. Results of electronic structure calculations for URhGe[63] suggest itinerant *5f*-electron ferromagnetism also in this compound. However, the ferromagnetic order in URhGe is unusually stable in applied hydrostatic pressure. The Curie temperature increases with linearly increasing applied pressure up to ≈ 17.5 K at 13 GPa[64].

The increase of $T_C$ with pressure in metallic ferromagnets is usually associated with more localized magnetic states, i.e. 5*f*-electron states in U compounds. $UGa_2$ can be taken as a prominent example[65, 66]. Very different degrees of localization can be identified e.g. by the response of critical temperature to external pressure. It has been demonstrated by an opposite pressure effect on $T_C$ in the ferromagnets, namely a negative effect in the case of UCoGa with more delocalized 5*f*-states and a positive effect in URhGa having the 5*f*-states considerably more localized leading to much larger U magnetic moments than in the UCoGa case[64, 67], represents an analogy to the pair UCoGe, URhGe. In this scenario, the UCoGe is a weak itinerant 5*f*-electron ferromagnet with the spontaneous moment of 0.07 $\mu_B$/f.u. and URhGe has considerably more localized 5*f*-electron states yielding a stable U moment of 0.43 $\mu_B$/f.u. This is also well documented by the stability of ferromagnetism in $URh_{1-x}Ir_xGe$ which persists up to $x$ = 0.43 whereas the ferromagnetism in $UCo_{1-x}Ir_xGe$ ceases with only slight Ir doping of $x < 0.02$. A tiny U magnetic moment in UCoGe appears on the verge of instability. It is known that no ferromagnetic order could be observed in some UCoGe samples[50, 68]. Therefore the substitutional disorder in $UCo_{0.98}Ir_{0.02}Ge$ could be considered as the mechanism suppressing the ordered U moment[67, 69, 70].

Important arguments corroborating the very different degree of localization of 5*f*-electrons and consequently the different stability of ferromagnetism in UCoGe and URhGe provide the values of lattice parameters listed in Table. S7[32]. The URhGe unit cell is apparently in all three dimensions larger than that of UCoGe. This implies also a larger value of $d_{U\text{-}U}$ and other interatomic distances leading to a smaller 5*f*-5*f* overlap and weaker 5*f*-ligand hybridization, which results in a less reduced U magnetic moment in URhGe. UIrGe and URhGe have very similar lattice parameters arising from almost identical radii of the transition element ions[71]. Thus, the changes in lattice parameters due to substitutions in the $URh_{1-x}Ir_xGe$ system are almost negligible. Then stable FM and AFM phases on the Rh and Ir side, respectively, of the phase diagram can be expected.

## CONCLUSIONS

This work intended to explore the evolution of magnetism in the pseudo-ternary compounds of composition between the superconducting itinerant 5*f*-electron ferromagnet UCoGe and isostructural and isoelectronic antiferromagnet UIrGe. For this purpose, we have prepared a series of polycrystalline samples of UCoGe doped with Ir in the wide range of concentrations $UCo_{1-x}Ir_xGe$ ($0.005 \le x \le 0.9$) and ten representative single crystals, and studied them by measuring magnetization, specific heat, and electrical resistivity at various temperatures and magnetic fields. The rich data set enabled us to construct the complex *T*-*x* phase diagram of the entire pseudo-ternary system.

The already very low doping of UCoGe by Ir ($x$ = 0.02) leads to the instant suppression of ferromagnetism and superconductivity.

Further increase of the Ir content reinforces the magnetic correlations resulting in a strange state in the interval $0.24 \le x \le 0.8$. In our view, we ascribe magnetic behavior in this concentration range to the paramagnetic state with strong FM/AFM correlations (a correlated paramagnet – (CPM)) which can lead to the freezing of incoherent spin arrangements (spin-glass-like, cluster glass [49] with possible short-range magnetic order (SRMO)) at low temperatures. Ordinary antiferromagnetism has been detected in the interval $0.84 \le x \le 1$ with an abrupt increase of $T_N$ towards the parent UIrGe. The character of correlations depends on the evolution of U magnetic moments and the hierarchy of FM and AFM exchange interactions mediated by 5*f*-ligand hybridization 5*f*-3*d* / 5*d* (U-Co / Ir), which is controlled by the composition of the Co/Ir sublattice.

Our present work in comparison with[15] experimentally demonstrates the fundamentally different transformation between the FM and AFM states of parent compounds in $URh_{1-x}Ir_xGe$ and $UCo_{1-x}Ir_xGe$ system, respectively. The striking difference is most probably due to the considerably different degree of the localization of U 5*f*-electron states in the weak itinerant ferromagnet UCoGe with tiny U magnetic moment on the verge of instability and the rather stable U moment in URhGe characterized by more localized 5*f* states.

The polycrystalline study has revealed basic knowledge about the *T*-*x* phase diagram of the $UCo_{1-x}Ir_xGe$ system, but similarly, as in the neighboring alloy systems, the complete single-crystal study allowed us to understand the system in more detail. Results of the detailed study of magnetization of single crystals in the main crystallographic directions demonstrate the gradual transformation of magnetocrystalline anisotropy from the uniaxial anisotropy of the Ising-like magnetism in UCoGe to the orthorhombic anisotropy in UIrGe.

The temperatures of the *b*- and *c*-axis maxima characteristic for CPM regime have been determined existing above the different magnetic states and therefore the correlations can be of various natures. Further rigorous investigation and understanding of the competition between the FM and AFM correlations in the $UCo_{1-x}Ir_xGe$ system could be useful particularly in the frame of the detected CPM region in the recently discovered heavy-fermion superconductor $UTe_2$.

**Acknowledgments**

The experiments were performed at MGML (http://mgml.eu), which is supported within the program of Czech Research Infrastructures (project no. LM2018096). This project was supported by OP VVV project MATFUN under Grant No.CZ.02.1.01/0.0/0.0/15_003/0000487. This project was supported by OP VVV project MATFUN under Grant No. CZ.02.1.01/0.0/0.0/15_003/0000487. This project was supported by VEGA 1/0404/21, 1/0705/20 and APVV-16-0079 and also project No. 001PU-2-1/2018. The authors are indebted to Dr. Ross Colman for critical reading and correcting the manuscript.

REFERENCES

[1] T. V. Bay, A. M. Nikitin, T. Naka, A. McCollam, Y. K. Huang, and A. de Visser, Physical Review B **89**, 214512 (2014).
[2] N. T. Huy, A. Gasparini, D. E. de Nijs, Y. Huang, J. C. P. Klaasse, T. Gortenmulder, A. de Visser, A. Hamann, T. Gorlach, and H. von Löhneysen, Physical Review Letters **99**, 067006 (2007).
[3] D. Aoki and J. Flouquet, Journal of the Physical Society of Japan **81**, 011003 (2012).
[4] M. Vališka, J. Pospíšil, M. Diviš, J. Prokleška, V. Sechovský, and M. M. Abd-Elmeguid, Physical Review B **92**, 045114 (2015).
[5] J. Pospíšil, J. Gouchi, Y. Haga, F. Honda, Y. Uwatoko, N. Tateiwa, S. Kambe, S. Nagasaki, Y. Homma, and E. Yamamoto, Journal of the Physical Society of Japan **86**, 044709 (2017).
[6] D. Aoki, A. Huxley, E. Ressouche, D. Braithwaite, J. Flouquet, J. P. Brison, E. Lhotel, and C. Paulsen, Nature **413**, 613 (2001).
[7] J. Pospíšil, Y. Haga, Y. Kohama, A. Miyake, S. Kambe, N. Tateiwa, M. Vališka, P. Proschek, J. Prokleška, V. Sechovský, et al., Physical Review B **98**, 014430 (2018).
[8] S. Yoshii, A. V. Andreev, E. Brück, J. C. P. Klaasse, K. Prokeš, F. R. d. Boer, M. Hagiwara, K. Kindo, and V. Sechovský, Journal of Physics: Conference Series **51**, 151 (2006).
[9] S. Kawamata, K. Ishimoto, Y. Yamaguchi, and T. Komatsubara, Journal of Magnetism and Magnetic Materials **104-107**, 51 (1992).
[10] F. R. de Boer, K. Prokes, H. Nakotte, E. Brück, M. Hilbers, P. Svoboda, V. Sechovsky, L. Havela,

and H. Maletta, Physica B **201**, 251 (1994).
[11] V. Sechovsky, L. Havela, A. Purwanto, A. C. Larson, R. A. Robinson, K. Prokes, H. Nakotte, E. Brück, F. R. de Boer, P. Svoboda, et al., Journal of Alloys and Compounds **213-214**, 536 (1994).
[12] V. H. Tran, R. Tro†, and D. urski, Journal of Magnetism and Magnetic Materials **87**, 291 (1990).
[13] R. Tro† and V. H. Tran, Journal of Magnetism and Magnetic Materials **73**, 389 (1988).
[14] W. Knafo, T. D. Matsuda, F. Hardy, D. Aoki, and J. Flouquet, Physical Review B **100**, 094421 (2019).
[15] J. Pospíšil, Y. Haga, S. Kambe, Y. Tokunaga, N. Tateiwa, D. Aoki, F. Honda, A. Nakamura, Y. Homma, E. Yamamoto, et al., Physical Review B **95**, 155138 (2017).
[16] C. Duan, K. Sasmal, M. B. Maple, A. Podlesnyak, J.-X. Zhu, Q. Si, and P. Dai, Physical Review Letters **125**, 237003 (2020).
[17] S. Sundar, S. Gheidi, K. Akintola, A. M. Côté, S. R. Dunsiger, S. Ran, N. P. Butch, S. R. Saha, J. Paglione, and J. E. Sonier, Physical Review B **100**, 140502 (2019).
[18] O. Eriksson, M. Brooks, B. Johansson, R. Albers, and A. Boring, Journal of applied physics **69**, 5897 (1991).
[19] J. L. Smith and E. A. Kmetko, Journal of the Less Common Metals **90**, 83 (1983).
[20] B. Cooper, R. Siemann, D. Yang, P. Thayamballi, and A. Banerjea, in *Handbook on the physics and chemistry of the actinides. Vol. 2*, 1985).
[21] G.-J. Hu, N. Kioussis, A. Banerjea, and B. R. Cooper, Physical Review B **38**, 2639 (1988).
[22] V. Sechovsky and L. Havela, in *Handbook of Magnetic Materials* (Elsevier, 1998), Vol. 11, p. 1.
[23] K. Prokeš, T. Tahara, Y. Echizen, T. Takabatake, T. Fujita, I. H. Hagmusa, J. C. P. Klaasse, E. Brück, F. R. de Boer, M. Diviš, et al., Physica B: Condensed Matter **311**, 220 (2002).
[24] S. El-Khatib, S. Chang, H. Nakotte, D. Brown, E. Brück, A. J. Schultz, A. Christianson, and A. Lacerda, Journal of Applied Physics **93**, 8352 (2003).
[25] K. Prokes, T. Tahara, Y. Echizen, T. Takabatake, T. Fujita, I. H. Hagmusa, J. Bruck, F. R. de Boer, M. Divis, and V. Sechovsky, Physica B-Condensed Matter **334**, 272 (2003).
[26] M. O. Steinitz, E. Fawcett, C. E. Burleson, J. A. Schaefer, L. O. Frishman, and J. A. Marcus, Physical Review B **5**, 3675 (1972).
[27] S. Kawamata, K. Ishimoto, Y. Yamaguchi, and T. Komatsubara, Journal of Magnetism and Magnetic Materials **104**, 51 (1992).
[28] J. Pospíšil, Y. Haga, A. Miyake, S. Kambe, Y. Tokunaga, M. Tokunaga, E. Yamamoto, P. Proschek, J. Volný, and V. Sechovský, Physical Review B **102**, 024442 (2020).
[29] N. T. Huy, D. E. de Nijs, Y. K. Huang, and A. de Visser, Physical Review Letters **100**, 077002 (2008).
[30] J. Pospíšil, Y. Haga, A. Miyake, S. Kambe, N. Tateiwa, Y. Tokunaga, F. Honda, A. Nakamura, Y. Homma, M. Tokunaga, et al., Physica B: Condensed Matter **536**, 532 (2018).
[31] M. Valiska, J. Pospisil, G. Nenert, A. Stunault, K. Prokes, and V. Sechovsky, JPS Conf. Proc. **3**, 012011 (2014).
[32] Supplementary
[33] A. Gasparini, Y. K. Huang, N. T. Huy, J. C. P. Klaasse, T. Naka, E. Slooten, and A. de Visser, Journal of Low Temperature Physics **161**, 134 (2010).
[34] P. F. de Châtel, K. Prokeš, H. Nakotte, A. Purwanto, V. Sechovský, L. Havela, E. Brück, R. A. Robinson, and F. R. de Boer, Journal of Magnetism and Magnetic Materials **177-181**, 785 (1998).
[35] K. Prokeš, P. De Châtel, E. Brück, F. De Boer, K. Ayuel, H. Nakotte, and V. Sechovský, Physical Review B **65**, 144429 (2002).
[36] K. Prokeš, H. Nakotte, E. Brück, P. De Chatel, and V. Sechovský, Applied Physics A **74**, s757 (2002).
[37] M. J. Besnus, J. P. Kappler, P. Lehmann, and A. Meyer, Solid State Communications **55**, 779

(1985).
[38] K. Prokeš, T. Tahara, T. Fujita, H. Goshima, T. Takabatake, M. Mihalik, A. A. Menovsky, S. Fukuda, and J. Sakurai, Physical Review B **60**, 9532 (1999).
[39] A. Gasparini, Y. K. Huang, J. Hartbaum, H. von Löhneysen, and A. de Visser, Physical Review B **82**, 052502 (2010).
[40] J. Pospíšil, J. Gouchi, Y. Haga, F. Honda, Y. Uwatoko, N. Tateiwa, S. Kambe, S. Nagasaki, Y. Homma, and E. Yamamoto, Journal of the Physical Society of Japan **86**, 044709 (2017).
[41] R. Elliott and F. Wedgwood, Proceedings of the Physical Society (1958-1967) **81**, 846 (1963).
[42] D. McWhan and d. T. Rice, Physical Review Letters **19**, 846 (1967).
[43] B. Stebler, Physica Scripta **2**, 53 (1970).
[44] K. Prokeš, H. Nakotte, V. Sechovský, M. Mihalik, and A. V. Andreev, Physica B: Condensed Matter **350**, E199 (2004).
[45] K. Prokeš, V. Sechovský, F. R. de Boer, and A. V. Andreev, Journal of Physics: Condensed Matter **20**, 104221 (2008).
[46] H. Kadowaki, Journal of Physics and Chemistry of Solids **60**, 1199 (1999).
[47] B. Janoušová, P. Svoboda, V. Sechovský, K. Prokeš, T. Komatsubara, H. Nakotte, S. Chang, B. Ouladdiaf, and I. Císařová, Applied Physics A **74**, s731 (2002).
[48] D. Gralak, A. J. Zaleski, and V. H. Tran, Journal of Solid State Chemistry **242**, 175 (2016).
[49] S. Pakhira, N. S. Sangeetha, V. Smetana, A. V. Mudring, and D. C. Johnston, Journal of Physics: Condensed Matter **33**, 115802 (2020).
[50] J. Pospíšil, K. Prokeš, M. Reehuis, M. Tovar, J. P. Vejpravová, J. Prokleška, and V. Sechovský, Journal of the Physical Society of Japan **80**, 084709 (2011).
[51] W. Knafo, R. Settai, D. Braithwaite, S. Kurahashi, D. Aoki, and J. Flouquet, Physical Review B **95**, 014411 (2017).
[52] D. Aoki, W. Knafo, and I. Sheikin, Comptes Rendus Physique **14**, 53 (2013).
[53] W. Knafo, T. D. Matsuda, D. Aoki, F. Hardy, G. W. Scheerer, G. Ballon, M. Nardone, A. Zitouni, C. Meingast, and J. Flouquet, Physical Review B **86**, 184416 (2012).
[54] D. D. Koelling, B. D. Dunlap, and G. W. Crabtree, Physical Review B **31**, 4966 (1985).
[55] H. H. Hill, *Plutonium 1970 and Other Actinides* ( Edited by W. N. Miner (American Institute of Mining, Metallurgical, and Petroleum Engineers, New York). , 1970).
[56] J. M. Fournier, A. Boeuf, P. Frings, M. Bonnet, J. v. Boucherle, A. Delapalme, and A. Menovsky, Journal of the Less Common Metals **121**, 249 (1986).
[57] L. Severin, L. Nordström, M. S. S. Brooks, and B. Johansson, Physical Review B **44**, 9392 (1991).
[58] M. W. Butchers, J. A. Duffy, J. W. Taylor, S. R. Giblin, S. B. Dugdale, C. Stock, P. H. Tobash, E. D. Bauer, and C. Paulsen, Physical Review B **92**, 121107 (2015).
[59] M. Taupin, L.-P. Sanchez, J. P. Brison, D. Aoki, G. Lapertot, F. Wilhelm, and A. Rogalev, Physical Review B **92**, 035124 (2015).
[60] E. Hassinger, D. Aoki, G. Knebel, and J. Flouquet, Journal of the Physical Society of Japan **77**, 073703 (2008).
[61] E. Slooten, T. Naka, A. Gasparini, Y. K. Huang, and A. de Visser, Physical Review Letters **103**, 097003 (2009).
[62] G. Bastien, D. Braithwaite, D. Aoki, G. Knebel, and J. Flouquet, Physical Review B **94**, 125110 (2016).
[63] M. Divis, P. Mohn, K. Schwarz, P. Blaha, and P. Novak, in *1st International Workshop on Electron Correlations and Materials Properties*, Iraklion, Greece, 1998), p. 487.
[64] F. Hardy, A. Huxley, J. Flouquet, B. Salce, G. Knebel, D. Braithwaite, D. Aoki, M. Uhlarz, and C. Pfleiderer, Physica B: Condensed Matter **359-361**, 1111 (2005).
[65] A. Kolomiets, J.-C. Griveau, J. Prchal, A. Andreev, and L. Havela, Physical Review B **91**, 064405 (2015).

[66] B. Chatterjee and J. Kolorenč, Physical Review B **103**, 205146 (2021).
[67] P. Opletal, J. Valenta, P. Proschek, V. Sechovský, and J. Prokleška, Physical Review B **102**, 094409 (2020).
[68] J. Vejpravova-Poltierova, J. Pospisil, J. Prokleska, K. Prokes, A. Stunault, and V. Sechovsky, Physical Review B **82**, 180517 (2010).
[69] M. Míšek, J. Prokleška, P. Opletal, P. Proschek, J. Kaštil, J. Kamarád, and V. Sechovský, AIP Advances **7**, 055712 (2017).
[70] M. Míšek, P. Proschek, P. Opletal, V. Sechovský, J. Kaštil, J. Kamarád, M. Žáček, and J. Prokleška, AIP Advances **8**, 101316 (2018).
[71] B. Cordero, V. Gómez, A. E. Platero-Prats, M. Revés, J. Echeverría, E. Cremades, F. Barragán, and S. Alvarez, Dalton Transactions, 2832 (2008).

Supplemental Material for:

# Alloying driven transition between ferro- and antiferromagnetism in UTGe compounds: the $UCo_{1-x}Ir_xGe$ case

Dávid Hovančík[1,2,], Akinari Koriki[1,3], Anežka Bendová[1], Petr Doležal[1], Petr Proschek[1], Martin Míšek[4], Marian Reiffers[2], Jan Prokleška[1], Jiří Pospíšil[1] and Vladimír Sechovský[1]

[1]*Charles University, Faculty of Mathematics and Physics, Department of Condensed Matter Physics, Ke Karlovu 5, 121 16 Prague 2, Czech Republic*
[2]*University of Presov, Faculty of Humanities and Natural Sciences, 17 Novembra 1, 081 16 Presov, Slovakia*
[3]*Hokkaido University, Graduate School of Science, Department of Condensed Matter Physics, Kita10, Nishi 8, Kita-ku, Sapporo, 060-0810, Japan*
[4]*Institute of Physics, Academy of Sciences of Czech Republic, v.v.i, Na Slovance 2, 182 21 Prague 8, Czech Republic*

## EXPERIMENTAL

To study the development of the magnetic states in the $UCo_{1-x}Ir_xGe$ system we have prepared a series of polycrystalline samples with different Co/Ir concentrations to fully cover all regions of expected interesting magnetic property changes. All samples were prepared by arc melting from the stoichiometric amounts of the elements (purity of Co 4N5, Ge 6N, and Ir 4N). U was purified by the solid-state electrotransport technique, following our previous experience with the preparation of UCoGe[49]. The arc-melting process was realized under a protective Ar (6N purity) atmosphere on a water-cooled Cu crucible. Each sample was three times turned upside down and subsequently re-melted to achieve the best homogeneity. All samples were separately wrapped into a Ta foil (99.99%), sealed in a quartz tube under a vacuum of ~$10^{-6}$ mbar, annealed at 900 °C for 7 days, and then slowly cooled down to room temperature to avoid the creation of internal stresses. Each sample was characterized by X-ray powder diffraction (XRPD) at room temperature on a Bruker D8 Advance diffractometer. The obtained data were evaluated by the Rietveld technique using FULLPROF/WINPLOTR software[71], given the previously published crystallographic data of the isostructural UCoGe[2] and UIrGe[15] compounds as starting models.

Several single crystals were grown by Czochralski pulling for the melt of nominal compositions (see Table S1) in a tri-arc furnace using pulling speeds of 6 mm/h. The single crystals were several centimeters long cylinders with a diameter of 2-3 mm. The pulled crystals were wrapped in Ta foil, sealed in a quartz tube under the vacuum $10^{-6}$ mbar, and annealed for 14 days at 900°C. The quality of single crystals was verified by the Laue method and composition using the same analysis as polycrystalline samples. Oriented samples in the form of cubes for magnetization measurement were prepared using a fine abrasive wire saw.

The actual chemical composition (see Table S1[31]) was determined by a scanning electron microscope Tescan Mira I LMH equipped with an energy dispersive X-ray detector Bruker AXS. The specific heat ($C_p$) measurements were performed on thin polished plates using the relaxation method on a PPMS9T with a $^3$He insert. Magnetization ($M$) data were obtained on powder samples wrapped into the plastic ampoules using the VSM option of the PPMS9T and PPMS14T devices and the extraction method in an MPMS 7T magnetometer.

## Concentrations and lattice parameters

The XRPD analysis confirmed the orthorhombic TiNiSi-type structure (space group *Pnma*) of all samples over the entire concentration range in the $UCo_{1-x}Ir_xGe$ series. All three lattice parameters show a linear dependence on $x$ obeying Vegard's law see (Fig. S1[31])[72]. The increasing trend of the lattice parameters reflects a considerable difference between the covalent radii of the Co (126 pm) and Ir (141 pm) atoms [70]. The dependence of the lattice parameters and the distance between nearest-neighbor U ions ($d_{U-U}$) on $x$ is very similar to that of the $UCo_{1-x}Rh_xGe$ [27] presumably because of an almost identical transition elements radii (Rh ≈ Ir). Likewise, the value of the $d_{U-U}$ distance in both cases crosses Hill's critical value ($d_{U-U} \approx 3.5$ Å) at the same position $x \sim 0.6$.

Table S1: Nominal concentration of melt, corresponding real concentration determined by EDX, lattice parameters, and elementary lattice volume of studied single crystals of $UCo_{1-x}Rh_xGe$ compounds.

| Nominal concentration | Real concentration | *a* [Å] | *b* [Å] | *c* [Å] | V [Å$^3$] |
|---|---|---|---|---|---|
| $UCo_{0.99}Ir_{0.01}Ge$ | $UCo_{0.98}Ir_{0.02}Ge$ | 6.846 | 4.205 | 7.229 | 208.172 |
| $UCo_{0.95}Ir_{0.05}Ge$ | $UCo_{0.93}Ir_{0.07}Ge$ | 6.846 | 4.211 | 7.254 | 209.164 |
| $UCo_{0.84}Ir_{0.16}Ge$ | $UCo_{0.76}Ir_{0.24}Ge$ | 6.853 | 4.230 | 7.331 | 212.590 |
| $UCo_{0.73}Ir_{0.27}Ge$ | $UCo_{0.60}Ir_{0.40}Ge$ | 6.848 | 4.243 | 7.384 | 214.611 |
| $UCo_{0.50}Ir_{0.50}Ge$ | $UCo_{0.33}Ir_{0.67}Ge$ | 6.860 | 4.273 | 7.492 | 219.661 |
| $UCo_{0.25}Ir_{0.75}Ge$ | $UCo_{0.16}Ir_{0.84}Ge$ | 6.864 | 4.284 | 7.550 | 222.072 |
| $UCo_{0.25}Ir_{0.75}Ge$ | $UCo_{0.11}Ir_{0.89}Ge$ | 6.868 | 4.293 | 7.550 | 222.645 |
| $UCo_{0.15}Ir_{0.85}Ge$ | $UCo_{0.08}Ir_{0.92}Ge$ | 6.865 | 4.295 | 7.561 | 222.970 |

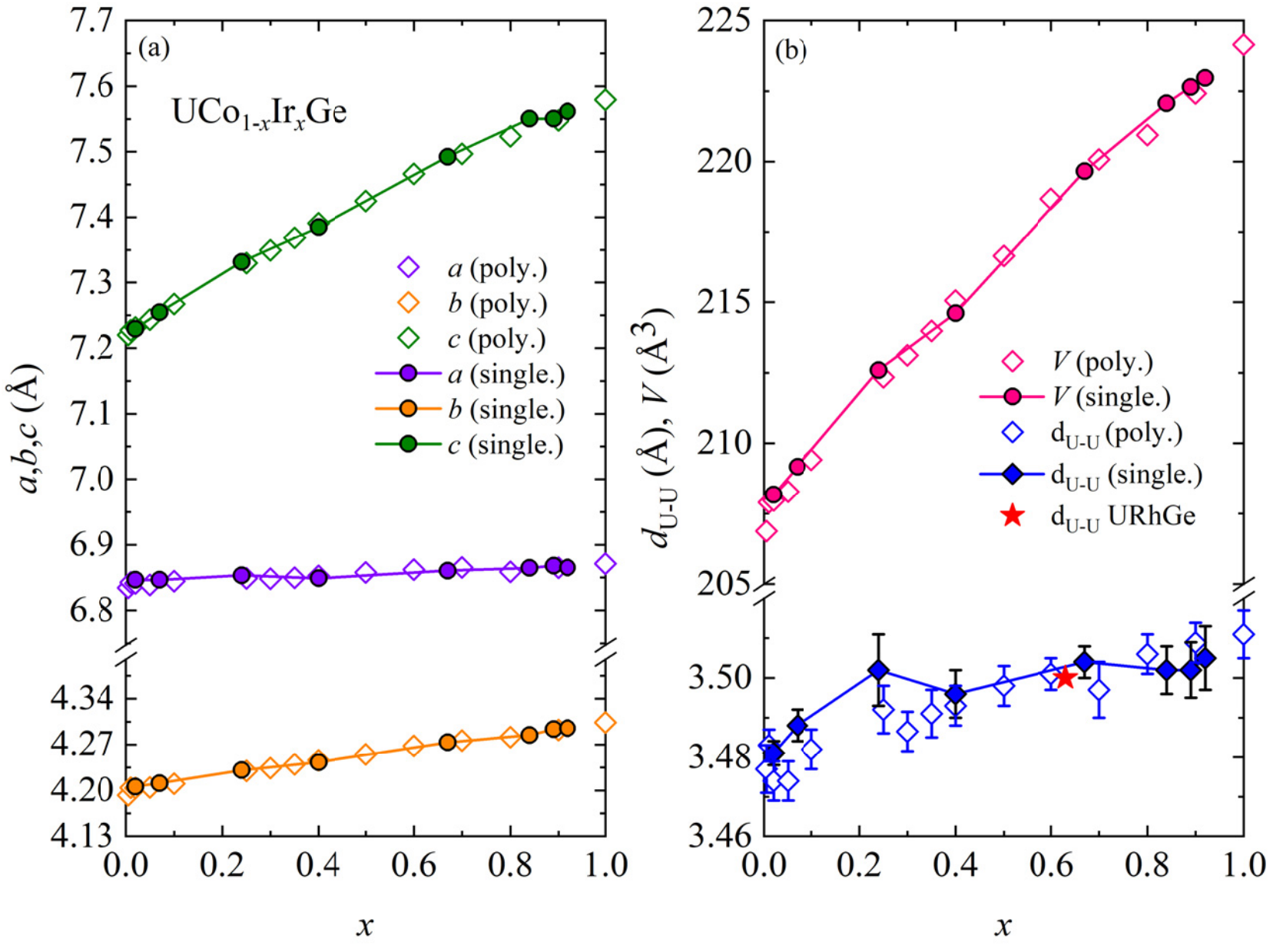

FIG. S1: Concentration dependence of (a) lattice parameters and (b) elementary lattice volume and $d_{U-U}$ distance in the poly and single-crystals of studied $UCo_{1-x}Ir_xGe$ compounds.

## Magnetization

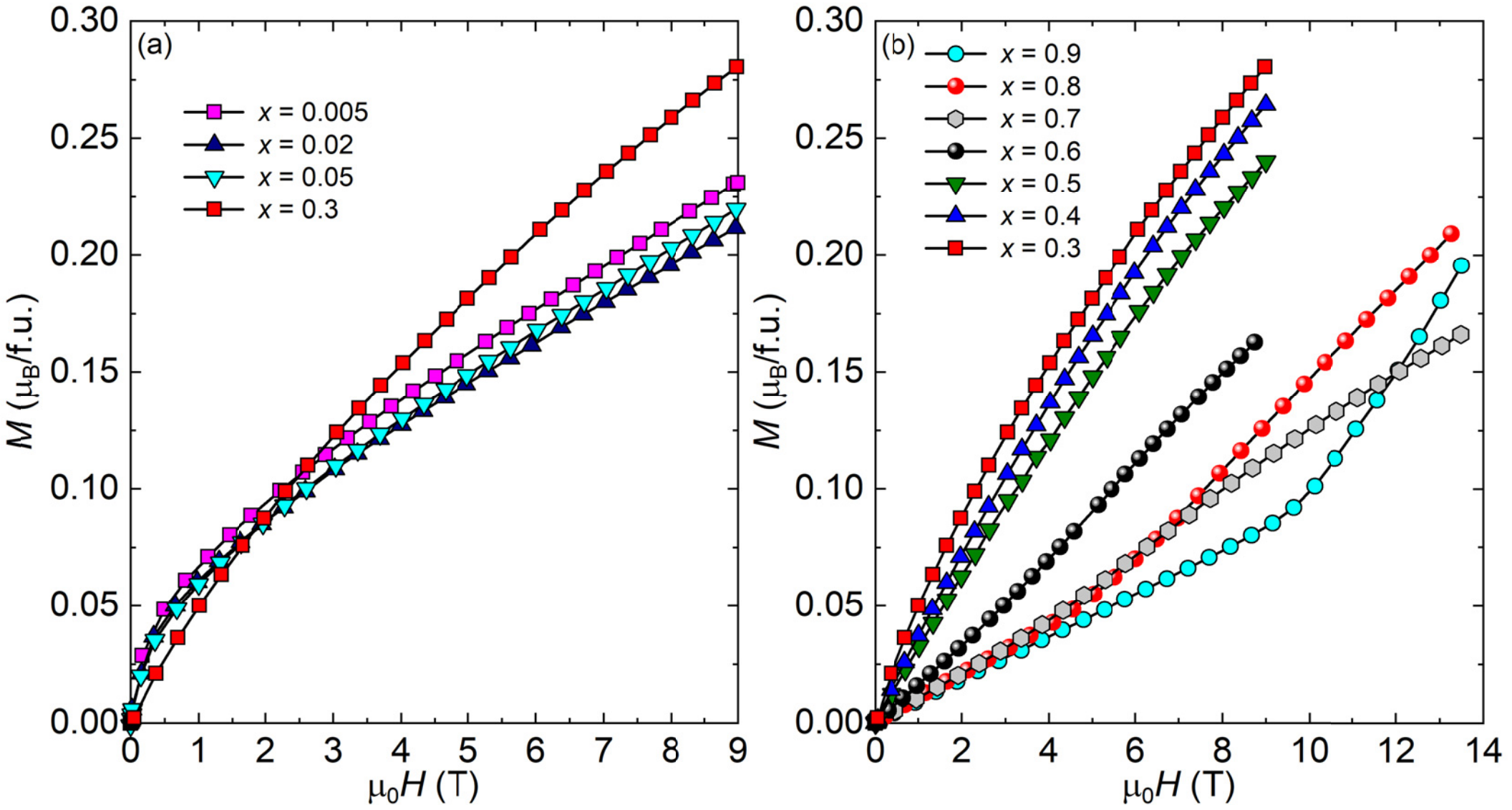

FIG. S2. The 2-K magnetization isotherms of selected $UCo_{1-x}Ir_xGe$ polycrystalline samples for (a) $x \leq 0.3$ and (b) $x \geq 0.4$.

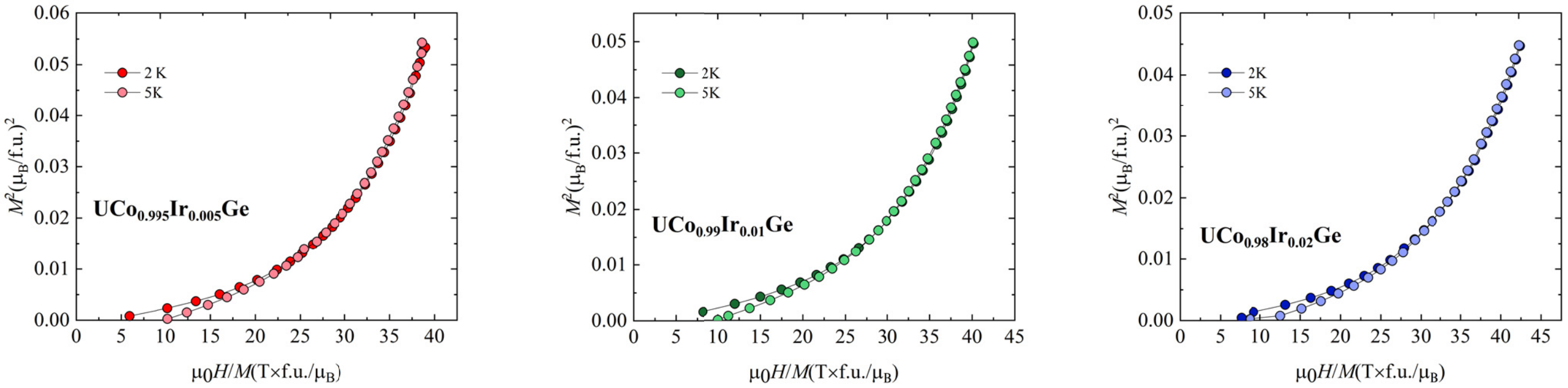

FIG. S3: Arrott plots of magnetization isotherm measured on $UCo_{1-x}Ir_xGe$ polycrystals with $x$ = a) 0.005, b) 0.01 and c) 0.02.

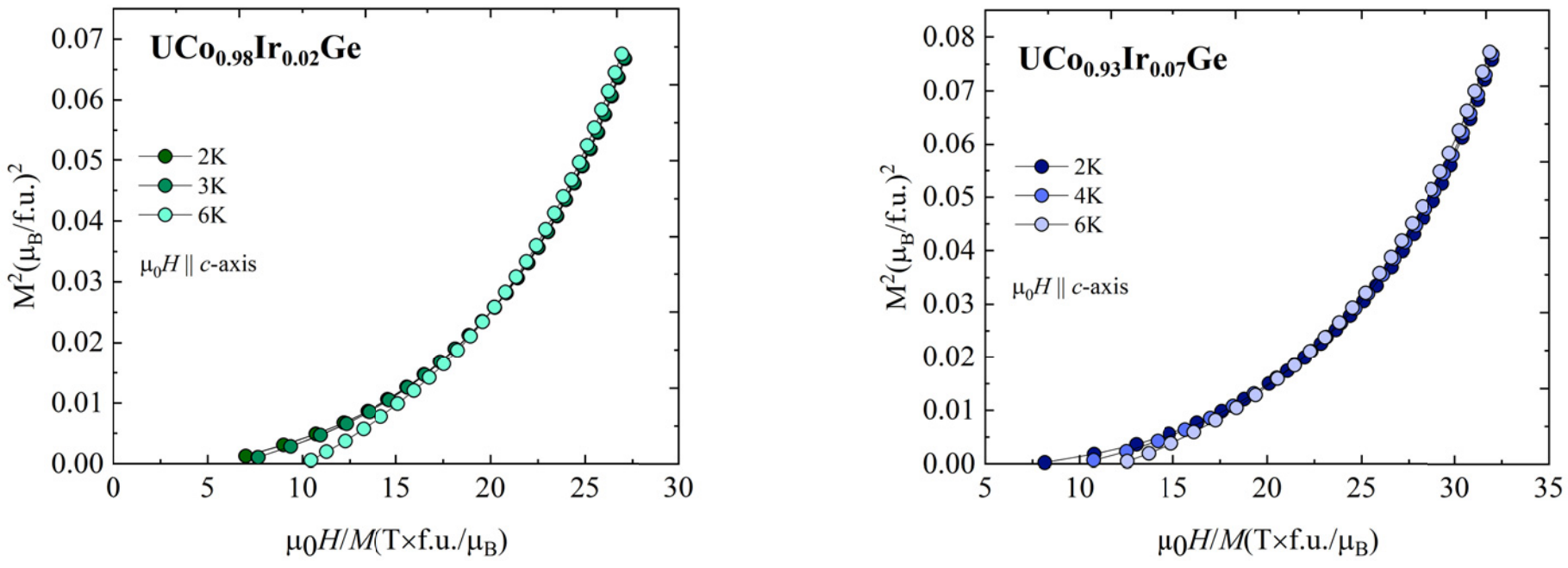

FIG. S4: Arrott plots of magnetization isotherm measured on $UCo_{1-x}Ir_xGe$ single crystals with $x$ = a)

0.02 and b) 0.07 in the magnetic field of 100 mT applied along the c-axis (easy magnetization direction).

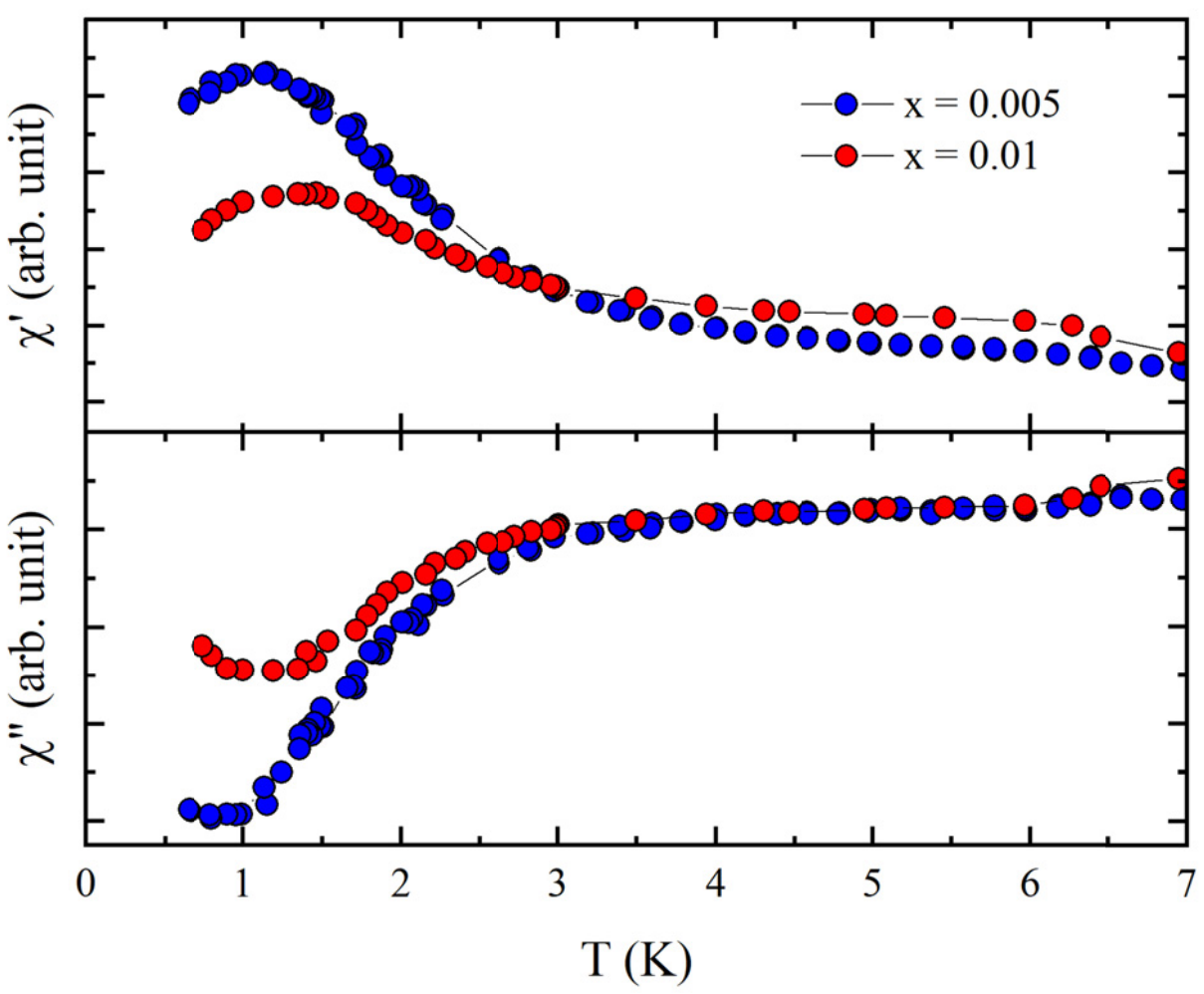

FIG. S5: Temperature dependences of the real ($\chi$') and imaginary ($\chi$'') component of the AC susceptibility of $UCo_{1-x}Ir_xGe$ polycrystals for $x$ = 0.005 and 0.01 at low temperatures.

Table S2: Characteristic parameters of 2-K magnetization isotherms of $UCo_{1-x}Rh_xGe$ single crystals in magnetic fields parallel to the *a*- and *b*-axis, respectively.

| Real Concentration | $M_{b\text{-}axis}$ ($\mu_0 H$ =14T) [$\mu_B$/f.u.] | d$M_{b\text{-}axis}$/d$\mu_0 H$ [$\mu_B$/T×f.u.] × $10^{-7}$ | $M_{a\text{-}axis}$ ($\mu_0 H$ =14T) [$\mu_B$/f.u.] | d$M_{a\text{-}axis}$/d$\mu_0 H$ [$\mu_B$/T×f.u.] × $10^{-7}$ |
|---|---|---|---|---|
| $UCo_{0.98}Ir_{0.02}Ge$ | 0.109 | 7.669 | 0.053 | 3.769 |
| $UCo_{0.93}Ir_{0.07}Ge$ | 0.121 | 8.592 | 0.042 | 3.051 |
| $UCo_{0.76}Ir_{0.24}Ge$ | 0.15 | 10.859 | 0.066 | 4.749 |
| $UCo_{0.60}Ir_{0.40}Ge$ | 0.147 | 10.5611 | 0.054 | 3.880 |

Table S3: Parameters of fitting the measured temperature dependences of paramagnetic susceptibility of $UCo_{1-x}Rh_xGe$ single crystals in magnetic fields parallel to the *c*- and *b*-axis, respectively.

| Real Concentration | $\chi_0(c)$ [$10^{-8}$ $m^3$/mol] | $\Theta(c)$ [K] | $\mu_{eff}(c)$ [$\mu_B$/f.u.] | $\chi_0(b)$ [$10^{-8}$ $m^3$/mol] | $\Theta(b)$ [K] | $\mu_{eff}(b)$ [$\mu_B$/f.u.] |
|---|---|---|---|---|---|---|
| $UCo_{0.98}Ir_{0.02}Ge$ | 0.886 | 2.768 | 2.080 | 0.547 | -164.338 | 2.736 |
| $UCo_{0.93}Ir_{0.07}Ge$ | 1.106 | 4.072 | 1.904 | 0.704 | -134.305 | 2.604 |
| $UCo_{0.76}Ir_{0.24}Ge$ | 1.256 | 7.701 | 1.847 | 0.592 | -104.802 | 2.611 |
| $UCo_{0.60}Ir_{0.40}Ge$ | 1.535 | 5.173 | 1.882 | 0.531 | -72.086 | 2.587 |
| $UCo_{0.33}Ir_{0.67}Ge$ | 1.180 | 0.057 | 1.724 | 0.423 | -39.070 | 2.508 |
| $UCo_{0.16}Ir_{0.84}Ge$ | 1.144 | -11.342 | 1.940 | 0.575 | -28.974 | 2.376 |
| $UCo_{0.11}Ir_{0.89}Ge$ | 1.117 | -8.554 | 1.642 | 1.102 | -39.563 | 2.350 |
| $UCo_{0.08}Ir_{0.92}Ge$ | 1.243 | -11.654 | 1.786 | 0.237 | -29.274 | 2.298 |
| UIrGe | 1.043 | -11.259 | 1.687 | 0.032 | -41.742 | 2.649 |

## Electrical resistivity

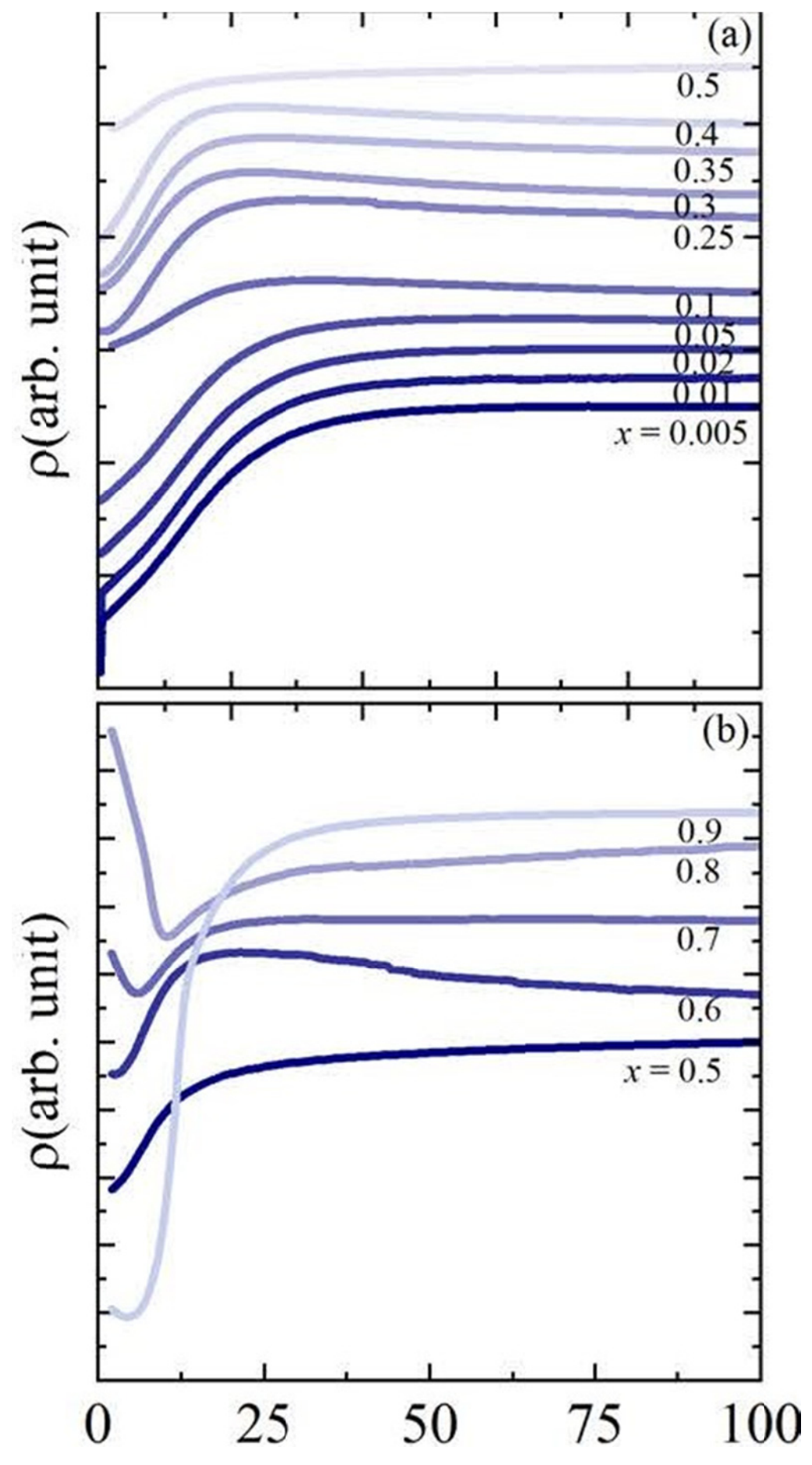

FIG. S6: Temperature dependences of relative electrical resistivity of $UCo_{1-x}Ir_xGe$ polycrystals (a) with $x \leq 0.5$, (b) with $x \geq 0.5$. The resistivity drop for x = 0.005 and 0.01 at $T < 0.5$ K is apparently due to the onset of superconductivity.

Table S4: The characteristic temperature $T(\rho_{max})$ of the $UCo_{1-x}Ir_xGe$ series as a function of $x$.

| $x$ | $T_{max}$ (K) |
|---|---|
| 0.10 | 32.6 |
| 0.25 | 31 |
| 0.30 | 23.5 |
| 0.35 | 26.2 |
| 0.40 | 23.1 |
| 0.60 | 23 |

Table S5: The values of the exponent $n$ defined in formula (2) determined from low-temperature $\rho(T)$ data measured on polycrystals of $UCo_{1-x}Ir_xGe$ compounds.

| $x$ | $n$ |
|---|---|
| 0.005 | 1.178 |
| 0.01 | 1.169 |
| 0.02 | 1.161 |
| 0.05 | 1.194 |
| 0.10 | 1.354 |
| 0.25 | 1.826 |
| 0.30 | 2.028 |
| 0.35 | 1.840 |
| 0.40 | 1.392 |
| 0.50 | .750 |

Table S6: The values of the exponent $n$ defined in formula (2) determined from low-temperature $\rho(T)$ data measured on single crystals of $UCo_{1-x}Ir_xGe$ compounds for current // *a-, b-* and *c*-axis, respectively.

| Real Concentration | $n(a)$ | $n$(b) | $n$(c) |
|---|---|---|---|
| $UCo_{0.98}Ir_{0.02}Ge$ | 1.291 | 1.156 | 0.688 |
| $UCo_{0.93}Ir_{0.07}Ge$ | 1.470 | 1.412 | 0.684 |
| $UCo_{0.76}Ir_{0.24}Ge$ | - | 1.257 | 1.468 |
| $UCo_{0.60}Ir_{0.40}Ge$ | 1.241 | 1.270 | - |

Table S7: Lattice parameters, unit cell volume and $d_{U\text{-}U}$ values for UCoGe, URhGe, and UIrGe.

| Compound | a(Å) | b(Å) | c(Å) | V(Å)3 | $d_{U-U}$(Å) | Reference |
|---|---|---|---|---|---|---|
| UCoGe | 6.8533(5) | 4.2098(3) | 7.2374(5) | 208.814 | 3.4808(4) | [27] |
| URhGe | 6.8902(11) | 4.3326(7) | 7.5184(14) | 224.34 | 3.5139(5) | [27] |
| UIrGe | 6.8714(4) | 4.3039(3) | 7.5793(5) | 224.15(6) | 3.511 | [15] |